\documentclass[manuscript]{acmart}
\usepackage[utf8]{inputenc}
\usepackage{soul,color}
\usepackage{tabularx}
\usepackage{booktabs}
\usepackage{multirow}
\usepackage{graphicx}
\usepackage{subcaption}
\usepackage{soul}

\AtBeginDocument{%
  }

\setcopyright{acmlicensed}
\copyrightyear{2018}
\acmYear{2018}
\acmDOI{XXXXXXX.XXXXXXX}

\acmConference[Conference acronym 'XX]{Make sure to enter the correct
  conference title from your rights confirmation email}{June 03--05,
  2018}{Woodstock, NY}

\acmISBN{978-1-4503-XXXX-X/2018/06}

\begin{document}

\title{Real-World Receptivity to Adaptive Mental Health Interventions: Findings from an In-the-Wild Study}

\author{Nilesh Kumar Sahu}
\email{nilesh21@iiserb.ac.in}
\orcid{0000-0003-1675-7270}
\affiliation{%
  \institution{Indian Institute of Science Education and Research Bhopal (IISERB)}
  \streetaddress{Bhauri}
  \city{Bhopal}
  \state{Madhya Pradesh}
  \country{India}
  \postcode{462066}
}

\author{Aditya Sneh}
\email{adityas19@iiserb.ac.in}
\affiliation{%
  \institution{Indian Institute of Science Education and Research Bhopal (IISERB)}
  \streetaddress{Bhauri}
  \city{Bhopal}
  \state{Madhya Pradesh}
  \country{India}
  \postcode{462066}
}

\author{Snehil Gupta}
\email{snehil.psy@aiimsbhopal.edu.in}
\orcid{0000-0001-5498-2917}
\affiliation{%
  \institution{All India Institute of Medical Sciences Bhopal}
  \city{Bhopal}
  \state{Madhya Pradesh}
  \country{India}
  \postcode{462026}
}

\author{Haroon R Lone}
\email{haroon@iiserb.ac.in}
\orcid{0000-0002-1245-2974}
\affiliation{%
  \institution{Indian Institute of Science Education and Research Bhopal (IISERB)}
  \streetaddress{Bhauri}
  \city{Bhopal}
  \state{Madhya Pradesh}
  \country{India}
  \postcode{462066}
}

\begin{abstract}

The rise of mobile health (mHealth) technologies has enabled real-time monitoring and intervention for mental health conditions using passively sensed smartphone data. Building on these capabilities, Just-in-Time Adaptive Interventions (JITAIs) seek to deliver personalized support at opportune moments, adapting to users' evolving contexts and needs.
Although prior research has examined how context affects user responses to generic notifications and general mHealth messages, relatively little work has explored its influence on engagement with actual mental health interventions. Furthermore, while much of the existing research has focused on detecting when users might benefit from an intervention, less attention has been paid to understanding \textit{receptivity} — users' willingness and ability to engage with and act upon the intervention.

In this study, we investigate user receptivity through two components: \textit{acceptance} (acknowledging or engaging with a prompt) and \textit{feasibility} (ability to act given situational constraints). We conducted a two-week in-the-wild study with 70 students using a custom Android app, LogMe, which collected passive sensor data and active context reports to prompt mental health interventions. The adaptive intervention module was built using Thompson Sampling, a reinforcement learning algorithm. We address four research questions relating smartphone features and self-reported contexts to acceptance and feasibility, and examine whether an adaptive reinforcement learning approach can optimize intervention delivery by maximizing a combined receptivity reward.  Our results show that several types of passively sensed data significantly influenced user receptivity to interventions. We found that the evening was the most feasible time to deliver interventions for higher acceptance. Participants were also more willing to perform suggested tasks when walking or eating.
Our findings contribute insights into the design of context-aware, adaptive interventions that are not only timely but also actionable in real-world settings.

\end{abstract}


\received{20 February 2007}
\received[revised]{12 March 2009}
\received[accepted]{5 June 2009}

\maketitle

\section{Introduction}

The advancement of mobile technologies has opened new opportunities in mobile health (mHealth) research to monitor mental health in real time. Passively sensed smartphone data can detect conditions such as stress \cite{mishra2020evaluating}, anxiety \cite{rashid2020predicting}, and mood changes \cite{meegahapola2023generalization}, providing valuable insights into individuals' daily lives \cite{englhardt2024classification}. Beyond monitoring, smartphones have enabled the delivery of interventions that offer timely support \cite{mishra2021detecting}, helping researchers observe and influence mental health as it unfolds naturally.

A recent approach known as Just-in-Time Adaptive Intervention (JITAI) has built upon these capabilities. JITAIs aim to deliver the right support at the right moment by personalizing interventions based on users' real-time context, symptom severity, and historical response patterns \cite{nahum2018just}. This framework adapts over time to maximize user engagement and intervention effectiveness across diverse settings and needs. A JITAI consists of two key components: the just-in-time aspect, which identifies opportune moments for delivering support, and the adaptive aspect, which tailors intervention content to the user’s current state or environment. An intervention is most effective when delivered at a moment that is both suitable and actionable \cite{mishra2021detecting, kunzler2019exploring}. For example, a breathing exercise intervention may be ignored if presented during work or in a busy social gathering, not because it was irrelevant, but because it was not feasible to act upon.

While most prior research has focused on the just-in-time delivery of interventions — such as detecting moments of stress or anxiety — relatively little work has examined participants' \textit{receptivity} to interventions based on their real-world context. Nahum-Shani et al. \cite{nahum2018just} defined \textit{receptivity} as a person’s ability to receive, process, and perform an intervention. In this framework, receiving refers to accepting the intervention notification by opening and viewing it, while performance refers to actually carrying out the suggested activity. Building on this framework, we define two key components of receptivity:
(1) \textbf{Acceptance}, meaning whether a participant accepts the intervention by acknowledging or engaging with the prompt; and (2) \textbf{Feasibility}, meaning whether the participant is able and willing to perform the intervention at that moment, considering social or environmental constraints. Performance may be affected if a participant dislikes the intervention or finds it infeasible to act upon, even if they initially accept it.

Within the mHealth and ubiquitous computing research communities, considerable attention has been given to understanding when individuals are most receptive to digital interventions \cite{kunzler2019exploring, pielot2015attention, koch2021drivers, liao2018just}. Two central concepts in this space are interruptibility, i.e., a person’s immediate readiness to respond to a notification, and engagement, which reflects the extent of interaction or sustained attention following that initial response \cite{pielot2015attention}. In the context of smartphone-based mental health support, engagement becomes especially crucial: it is not only about whether a user opens a notification but whether they meaningfully interact with its content. Prior work has largely focused on identifying optimal times or contexts (e.g., location or physical activity) to deliver general notifications, aiming to minimize disruption and maximize responsiveness \cite{fischer2010effects, morrison2017effect}. However, these strategies often fall short when applied to mental health interventions. Notifications that prompt behavioral actions, such as taking a mindfulness break or making a positive and negative list of one's behavior, differ significantly from routine alerts like messages or app updates. Unlike general notifications, which offer immediate utility or gratification, mental health support notifications may require cognitive effort, emotional readiness, or behavioral change. As a result, users may be more likely to ignore or postpone them, even when motivated to improve their well-being. This highlights a critical limitation in existing approaches: identifying moments of interruptibility alone is insufficient. For real-time mental health interventions to be truly effective, we must identify interruptible and contextually appropriate moments that are feasible for emotional engagement and action. Our work addresses this gap by exploring receptivity in smartphone-based mental health interventions in real-world settings.

To this end, we explored receptivity in terms of both acceptance and feasibility, where we used a combination of passively and actively sensed smartphone data from 70 participants. We conducted a two-week \textit{in-the-wild} study aimed at promoting mental health well-being among students. Using our custom-built Android application, \textit{LogMe}, participants received notifications prompting them to report their current activity context (e.g., studying, relaxing) and social context (e.g., being alone or with others), and were then presented with a suggested intervention to support their mental well-being.

Our study contributes to the field of ubiquitous computing by exploring how mobile sensing can inform the delivery of JITAIs. Specifically, we address the following research questions:

\begin{itemize} 
\item \textbf{RQ1:} How do passively sensed smartphone features (e.g., location, screen interaction) relate to the \textit{acceptance} of intervention notifications? 
\item \textbf{RQ2:} How do passively sensed smartphone features relate to the \textit{feasibility} of carrying out the suggested interventions? 
\item \textbf{RQ3:} How do participants' actively reported contextual factors (e.g., current activity, social setting) relate to the \textit{feasibility} of the suggested interventions? 
\item \textbf{RQ4:} Does an adaptive reinforcement learning algorithm improve intervention delivery by maximizing the \textit{average reward}, a metric reflecting the likelihood that an intervention is both accepted and feasible? \end{itemize}

Our exploratory analysis, guided by these research questions, revealed several meaningful patterns. For \textbf{RQ1}, we found that intervention acceptance was influenced by factors such as time of day, phone battery level, screen interaction, recognized activity, and location. The analysis of \textbf{RQ2} showed that intervention feasibility was similarly affected by daytime, battery status, screen use, app engagement, and location. In addressing \textbf{RQ3}, we observed that activities like walking and eating were associated with higher levels of both acceptability and feasibility. Finally, \textbf{RQ4} demonstrated that our adaptive intervention model consistently achieved a higher average reward across all user contexts, exceeding the baseline of 0.5. Collectively, these findings highlight key considerations for designing context-aware, adaptive interventions that are both timely and practical for real-world deployment.

\section{Related Work}

Recent Human sensing research has explored how smartphone apps can support student mental health through context-aware, personalized interventions \cite{kunzler2019exploring, mishra2020evaluating, liao2020personalized, saha2021person}. A prominent approach is the JITAI, which leverages sensor data (e.g., activity, location, time) to deliver timely support. For instance, a student’s phone might detect high stress before an exam and prompt a breathing exercise or motivational message. These systems aim to deliver “the right type and amount of support, at the right time” \cite{kunzler2019exploring, nahum2018just}. However, user receptivity varies based on personal (e.g., age, personality) and contextual (e.g., time of day, battery level) factors.

\subsection{Receptivity and Engagement Patterns}
Several studies have examined receptivity in student populations. Hamid et al. \cite{hamid2022you} developed a Cognitive Behavioral Therapy (CBT)-based app embedded in a storytelling game. Over several weeks, students reported improved mood and reflection, especially when narrative elements were gamified. However, the study was limited by a small sample and short follow-up. Bedmutha et al. \cite{bedmutha2024exploring} surveyed undergraduates about sharing passive data (e.g., location, phone use) and receiving mental health notifications. Students preferred tips and resources but were hesitant about sharing sensitive data or receiving alerts when busy. The study identified distinct user personas based on data-sharing preferences but was limited by self-reported data from a homogenous sample.

Timing and contextual relevance consistently emerge as key determinants of intervention success. Mishra et al. \cite{mishra2021detecting} found that context-triggered interventions achieved up to 40\% higher acceptance than randomly timed ones. Reviews like Nair et al.’s 4M model \cite{nair2021promoting} suggest that holistic interventions—addressing mindfulness, movement, meaning, and moderation—are more effective than single-strategy approaches.

\subsection{Intervention Types and Outcomes}
Empirical research supports the use of smartphone apps for delivering mental health interventions. Mello, a fully automated JITAI targeting repetitive negative thinking, significantly reduced depression and anxiety in a 6-week study (Cohen’s d = 0.5–0.87) \cite{bell2023personalized}. Users reported active self-management and high satisfaction. MindShift CBT, based on cognitive behavioral therapy, demonstrated feasibility in outpatient and inpatient settings for anxiety management \cite{sharma2022brief}. Similarly, MindScope used stress prediction models to deliver personalized feedback to students \cite{bogemann2023investigating}, though detailed results are still emerging. The DynaM-INT project \cite{abulfaraj2024impact} targets resilience in young adults through two mobile JITAIs that trigger interventions based on stress thresholds. This large-scale, longitudinal study aims to identify which users respond best to which strategies.

Several HCI studies have explored innovative delivery formats. MindShift \cite{wu2024mindshift} sent context-tailored persuasive messages to reduce smartphone overuse, achieving a 35\% acceptance rate—nearly triple that of random prompts. Effectiveness varied by emotional state (e.g., “comforting” messages were better during stress). Ally+, a JITAI for physical activity, used adaptive machine learning to time interventions, yielding 40\% higher engagement than random prompts. Other notable interventions include TypeOut \cite{xu2022typeout}, which replaced phone unlock screens with self-affirmation tasks, significantly reducing phone use, and Let’s Focus \cite{kim2017let}, which used geofencing to silence notifications during class. GeoAI JITAIs \cite{yang2019contextualizing} and BeActive \cite{choi2019multi} further demonstrate how location and cognitive state can inform multi-stage, context-driven prompting strategies.


\section{Methodology}

\subsection{Study Design and Data Collection}

The goal of this study was to investigate the \textit{receptivity} of adaptive interventions — specifically, participants' acceptance and feasibility — when real interventions were delivered in their natural, everyday environments. To support this goal, we developed LogMe, a custom-built smartphone application designed to capture users' activity and social contexts and to deliver contextually tailored interventions aimed at promoting mental health and well-being. The intervention content was developed in collaboration with clinical psychiatrists and psychologists to ensure relevance and appropriateness for the target population.

\begin{figure}[h]
    \centering
    \begin{subfigure}{0.3\textwidth}
        \includegraphics[width=\linewidth]{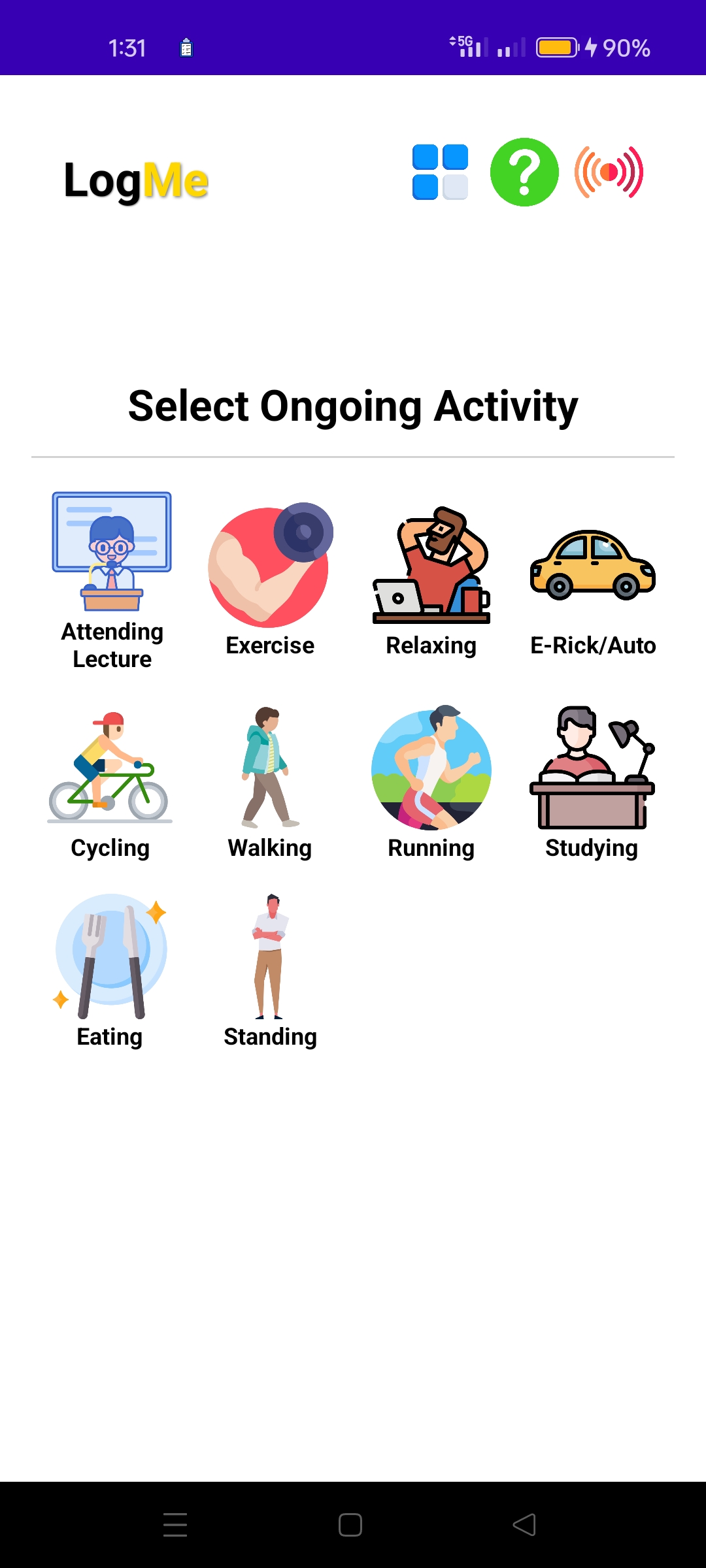}
        \Description{}
        \caption{Activity context selection}
        \label{fig:app_design_interface_1}
    \end{subfigure}
    \hfill
    \begin{subfigure}{0.3\textwidth}
        \includegraphics[width=\linewidth]{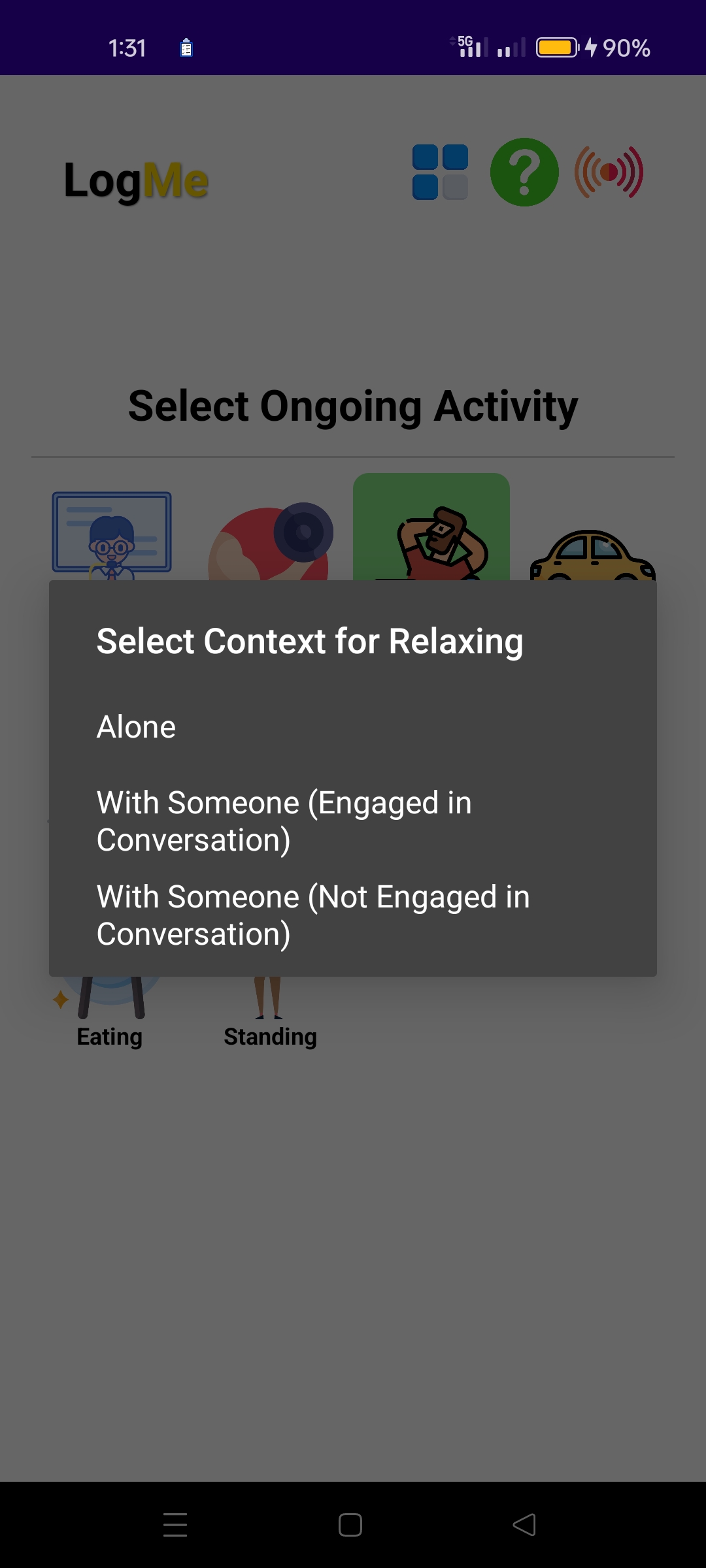}
        \Description{}
        \caption{Social context selection}
        \label{fig:app_design_interface_2}
    \end{subfigure}
    \hfill
    \begin{subfigure}{0.3\textwidth}
        \includegraphics[width=\linewidth]{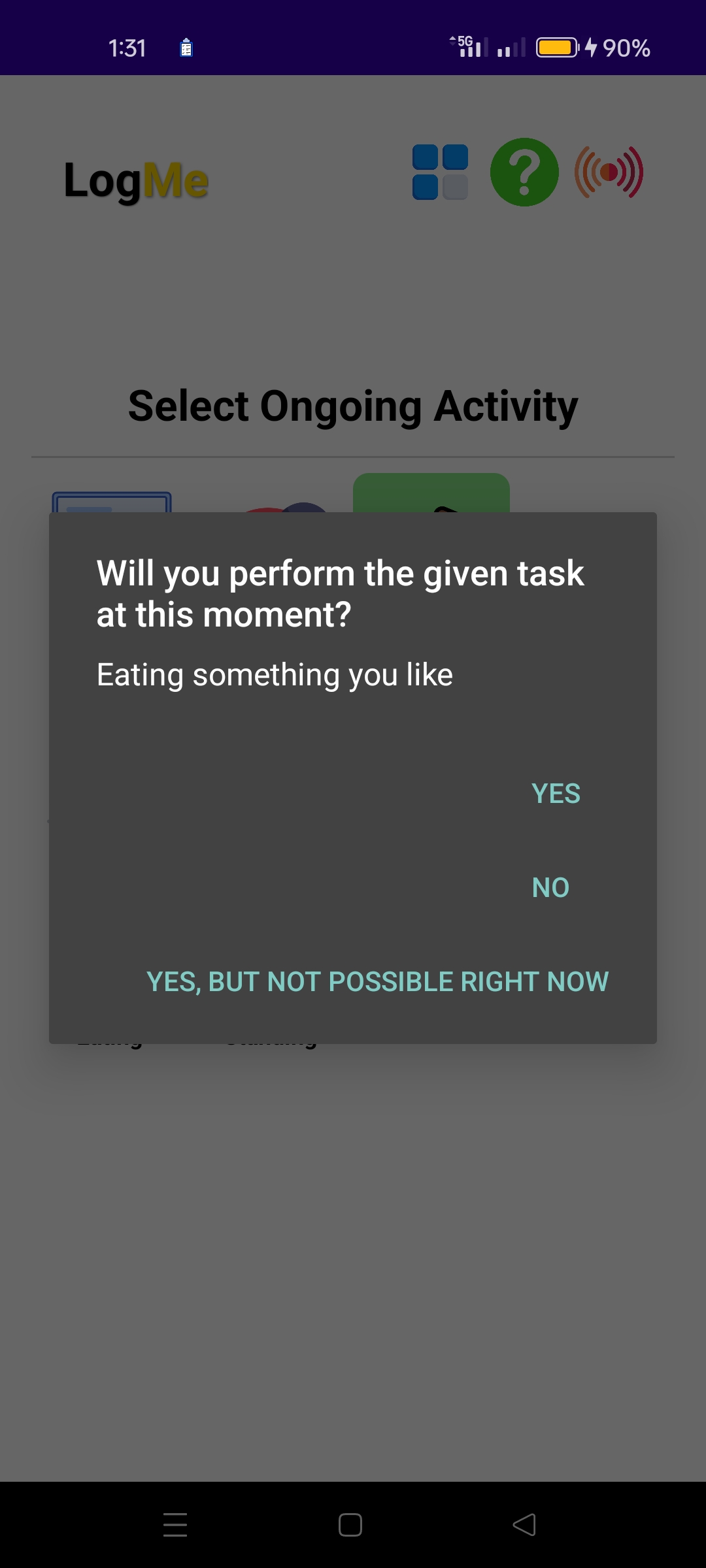}
        \Description{}
        \caption{Intervention Prompt}
        \label{fig:app_design_interface_3}
    \end{subfigure}
    \caption{Flow of user actions in the LogMe application following hourly notifications delivered between 7:00 AM and 8:00 PM.}
    \label{fig:app_design}
\end{figure}

LogMe is an Android application developed in Kotlin, designed to collect both passive sensor data and active user input on Android smartphones. It features users' context selection interface (see Figure \ref{fig:app_design_interface_1}). This interface is activated when users tap on hourly notifications. Upon selecting activity context, users are directed to a new screen where they choose their social context (see Figure \ref{fig:app_design_interface_2}). After both selections are made, the app’s backend intervention module recommends an appropriate intervention based on the chosen contexts (\ref{fig:app_design_interface_3}). The user interfaces for activity, social context selection, and intervention display are illustrated in Figure \ref{fig:app_design}.

\textbf{Active User data:} After receiving approval from the institutional ethics board, a general recruitment email was sent to the student community at the authors’ institution. Eighty interested students were invited for a briefing session for a two-week study and to install the LogMe study app. During the briefing, a research assistant explained the study procedures. Three participants chose not to continue due to privacy concerns. 

Following this, participants were shown a demonstration of the LogMe app on a dummy phone. They were guided through the app's interface and shown how the notifications would appear. Participants were informed that notifications for active data logging would be sent at the 55th minute of every hour, from 7:00 AM to 8:00 PM. A live demonstration illustrated what would happen after tapping a notification. 

Participants were then introduced to the activity context selection, in which they were asked to select the option that best described their current activity. The list of activity contexts was developed based on a prior survey conducted with the student population. From the survey results, we selected the top ten activity contexts that students typically engage in during their daily college life. These ten contexts were represented with images in the app interface (see Figure \ref{fig:app_design_interface_1}).

Next, participants were introduced to the concept of social context, with three options: (i) alone, (ii) with someone and engaged in conversation, and (iii) with someone but not engaged in conversation. Finally, the intervention interface was explained. The intervention interface consisted of a message: ``Will you perform the given task at this moment?'' along with a suggested task.  Participants were instructed to select one of three possible responses: (i) `Yes', (ii) `No', or (iii) `Yes, but not feasible right now'. It was clearly emphasized that participants were not required to actually perform the suggested task; they only needed to report their willingness. A response of `Yes' indicated willingness to perform the task; `No' indicated unwillingness due to personal preference or task aversion; and `Yes, but not feasible right' now indicated willingness but an inability to perform the task due to a social or environmental constraint.

\textbf{Passive sensor data:} The LogMe app collected passive data in the background, including screen interactions, battery status, phone calls, GPS location, app usage, and physical activity, using Google’s Activity Recognition API. Except for GPS and activity data, all other data types were event-driven, meaning they were logged only when a change occurred (e.g., call details were recorded only during incoming or outgoing calls). GPS data were sampled every 15 minutes, while activity data from the Google API were sampled every 30 seconds. Table \ref{tab:collected_sensor_data} provides an overview of the sensors and phone usage data collected, along with the type of information recorded from each source.

Three participants dropped out within 24 hours of app installation, and one withdrew within three days. Additionally, we excluded four participants who responded to fewer than 20 notifications. The final dataset included 70 participants (50 males and 20 females) with a mean age of 20.24 ($\sigma$=1.84).

\begin{table}[ht]
\small
\centering
\caption{Sensors and phone usage data collected in this study}
\label{tab:collected_sensor_data}
\begin{tabular}{p{2.3cm} p{9cm} p{2cm}}
\toprule
\textbf{} & \textbf{Data Collected} & \textbf{Sampling} \\
\midrule
App Usage & timestamp; app package name; foreground duration (ms); battery level (\%); screen brightness (0–255) & Event Based \\
Screen Interaction & timestamp; screen status (On/ Off); unlock event; screen-on duration (ms); off-period duration (ms) & Event Based \\
Battery & timestamp; status (Charging/ Full/ Discharging/ Not Charging); level (\%); voltage (mV); temperature (°C); power-saving mode & Event Based \\
Calls & timestamp; SHA-256(phone number); call type (Incoming/ Outgoing); start \& end times; duration (s); status & Event Based \\
GPS & timestamp; latitude, longitude, altitude (m) & Every 15 min \\
Google API Activity & timestamp; activity detected (e.g., still, walking, running); confidence (\%) & Every 30 s \\
\bottomrule
\end{tabular}
\end{table}

\subsection{Adaptive Intervention Design}
\subsubsection{Mental Wellbeing interventions}
A mental well-being intervention is a type of support designed to improve an individual's mental and emotional health \cite{sahu2018interdisciplinary}. These interventions aim to enhance coping skills, reduce stress, and promote positive mental health outcomes. Following the work of Barnett et al. \cite{barnett2021efficacy} and Worsley et al. \cite{worsley2020interventions}, and based on discussions with mental health professionals such as clinical psychiatrists, we identified four categories of interventions suitable for promoting mental well-being among students. These categories are: (1) Physical Activities (\#6 interventions), (2) Mental Relaxation (\#5 interventions), (3) Cognitive Activities (\#7 interventions), and (4) Emotional and Social Engagement (\#12 interventions). The Physical Activities category included interventions such as doing neck rolls and going for a short walk. The Mental Relaxation category included listening to music, closing one's eyes, and focusing on ambient sounds. Cognitive Activities involved counting backward from 100 to 1 or Making a list of positive things inside you. Emotional and Social Engagement included interventions like calling loved ones or recalling happy memories.

To ensure that the selected interventions were acceptable to the student community, we conducted a preference study with 40 participants (23 males and 17 females). Participants were asked: \textit{``Which of these tasks (i.e., interventions) would you prefer to perform during your daily activities to boost your mental well-being?''} Each participant selected four preferred interventions from each of the four categories. The final list of selected interventions used in the main study is presented in Table \ref{tab:intervention_prior_probalities}.



\subsubsection{Adaptive Mental Wellbeing interventions}

The JITAI has six key elements \cite{nahum2018just, kunzler2019exploring}: (1) the distal outcome, representing the long-term goal of the intervention; (2) proximal outcomes, which are short-term indicators serving as proxies for the distal outcome; (3) decision points, referring to specific times when the system evaluates whether to deliver an intervention; (4) intervention options, representing the range of possible actions that can be suggested; (5) tailoring variables, which are dynamic contextual or user-related factors used to guide the intervention selection; and (6) decision rules, which define how tailoring variables map to specific intervention options.

 In our study, we implemented the JITAI framework to assess the feasibility of delivering contextually relevant mental well-being interventions. The distal outcome was to determine whether our LogMe app could suggest interventions that participants are willing to perform within their current context. Proximal outcomes included the participant’s current activity that influenced the appropriateness of delivering an intervention at any given moment. The decision points were fixed, with intervention prompts scheduled at the 55th minute of each hour between 07:00 am and 08:00 pm, allowing for consistent delivery and evaluation across daily routines. As a tailoring variable, we used participants’ self-reported responses to the suggested interventions, specifically, whether they were willing to perform them. This approach aligned with our primary objective of evaluating intervention feasibility using an adaptive system. Based on these responses, the decision rule maps tailoring variables to specific intervention options, which is learned adaptively, enabling the system to adapt and learn over time.


In our study, we assumed that when a participant responded affirmatively to an intervention prompt, it indicated their willingness and ability to perform the suggested intervention, even though actual engagement was not tracked. No behavioral data were collected to confirm whether the intervention was carried out or whether it resulted in any psychological or behavioral effects. Participants were instructed to respond to each intervention prompt using one of the following three options:\textit{ `Yes' (reward = 1)}, \textit{`No' (reward = 0)}, and \textit{`Yes, but not feasible right now' (reward = 0.5)}. This response framework allowed us to capture both the perceived \textit{acceptability} and \textit{feasibility} of each intervention, offering valuable insights into how users evaluate suggested actions in real-time, real-world settings.


\subsubsection{Algorithm for Adaptive interventions}

Researchers have explored a variety of reinforcement learning (RL) algorithms to deliver adaptive interventions \cite{tewari2017ads}, including contextual multi-armed bandits (CMAB) \cite{huckvale2023protocol}, epsilon-greedy \cite{ismail2025improving}, upper confidence bound (UCB) \cite{rabbi2019optimizing}, Bayesian bandits \cite{liang2025context}, and Thompson Sampling \cite{kumar2024using}, etc. Selecting the appropriate RL model is crucial for maximizing the average reward, reflecting the suggested interventions' relevance and feasibility. Following the methodology of Kumar et al. \cite{kumar2024using}, we initially performed simulations using their publicly available code. We generated synthetic user responses based on different combinations of activity context, social context, and the selected list of interventions using Python's random function. We evaluated the performance of random selection, uniform selection, and several RL algorithms: epsilon-greedy, decaying epsilon, and Thompson Sampling. Consistent with the findings of Kumar et al. \cite{kumar2024using}, Thompson Sampling outperformed other algorithms in terms of average reward. Based on these results, we selected Thompson Sampling as the adaptive intervention algorithm and further evaluated its performance both with and without prior probabilities using simulation and human input. Here, the prior probabilities indicate how feasible each intervention is within a given context. In contrast, having no prior probabilities means that each intervention is initially assigned an equal probability within a given context (see Table \ref{tab:intervention_prior_probalities}, and Figure \ref{fig:app_design_interface_1}).

To estimate the prior feasibility probabilities of each intervention within different contexts, we conducted a separate survey with 19 students. Participants were asked whether the 16 shortlisted interventions (see Table \ref{tab:intervention_prior_probalities}) were feasible in each of the 10 activity contexts (i.e., attending lecture, exercise, relaxing, in vehicle, cycling, walking, running, studying, eating, standing). For every intervention-context pair, participants responded \textit{`Yes'} (feasible) or \textit{`No'} (not feasible). The initial probability was then computed as the proportion of `Yes' responses out of the total responses for that context. Intervention-context pairs with a feasibility probability lower than 0.4 were discarded, suggesting low feasibility. Additionally, to maintain exploration, intervention-context pairs with a probability of 1 were slightly reduced by 0.025 (chosen empirically) to ensure that other interventions could also be sampled during Thompson Sampling. This process revealed context-specific preferences—for example, no participants indicated willingness to perform "Journal Writing" during lectures, or tasks like "Writing down a worry and putting it aside" or "Making a list of positive things inside you" while walking. In contrast, in contexts like "Relaxing" and "Standing," all interventions were deemed feasible by participants. Table \ref{tab:intervention_prior_probalities} presents the obtained prior probabilities for each intervention across different contexts. Due to the varying number of feasible interventions per activity context, we developed separate adaptive intervention modules for each context, allowing for a personalized delivery of suggestions.

Subsequently, we re-ran Thompson Sampling simulations incorporating the collected prior probabilities. The results showed that including prior probabilities led to a higher average reward than the baseline with no priors. To further validate this, we developed a simple user interface where participants could select their current context and social context and then respond to the suggested interventions. Participants also reported a better experience when interventions were recommended using Thompson Sampling with prior probabilities, confirming the advantage of integrating prior knowledge into the adaptive system.

\begin{table}[]
\tiny
\caption{Prior probabilities of different interventions across various activity contexts. Probabilities marked with an asterisk * are below 0.4 and were excluded from implementation within that context. PA: Physical activity, MR: Mental relaxation, CA: Cognitive activities, ESE: Emotional and social engagement.}
\label{tab:intervention_prior_probalities}
\setlength{\tabcolsep}{4pt}
\begin{tabular}{@{}lcccccccccc@{}}
\toprule
\textbf{Intervention/ Context} & \textbf{Attending Lecture} & \textbf{Exercise} & \textbf{Relaxing} & \textbf{In vehicle} & \textbf{Cycling} & \textbf{Walking} & \textbf{Running} & \textbf{Studying} & \textbf{Eating} & \textbf{Standing} \\ \midrule
\textit{Breathing Exercise} \textbf{ (PA)} & 0.842& 0.842& 0.947& 0.526& 0.368*& 0.789& 0.474& 0.842& 0.158*& 0.972 \\
\textit{Calling or texting loved ones} \textbf{ (ESE)} & 0.211*& 0.579& 0.972& 0.972& 0.368*& 0.947& 0.316*& 0.579& 0.895& 0.972 \\
\textit{Close your eyes and try to hear ambient sounds} \textbf{ (MR)} & 0.526& 0.789& 0.972& 0.789& 0.474& 0.789& 0.421& 0.895& 0.526& 0.972 \\
\textit{Doing Simple Neck Rolls to Release Tension} \textbf{ (PA)} & 0.895& 0.972& 0.972& 0.895& 0.263*& 0.895& 0.316*& 0.92& 0.421& 0.972 \\
\textit{Eating something you like} \textbf{ (ESE)} & 0.158*& 0*& 0.972& 0.632& 0*& 0.474& 0*& 0.842& 0.972& 0.789 \\
\textit{Go for a small walk} \textbf{ (PA)} & 0.053*& 0.579& 0.737& 0*& 0*& 0.789& 0.368*& 0.895& 0.105*& 0.789 \\
\textit{Journal Writing} \textbf{ (CA)} & 0.368*& 0*& 0.789& 0.158*& 0*& 0.053*& 0.053*& 0.632& 0.053*& 0.421 \\
\textit{Listening to Music} \textbf{ (MR)} & 0.211*& 0.972& 0.972& 0.947& 0.895& 0.972& 0.867& 0.789& 0.842& 0.972 \\
\textit{Make a list of positive things inside you} \textbf{ (CA)} & 0.684& 0.421& 0.947& 0.474& 0.211*& 0.368*& 0.211*& 0.789& 0.368*& 0.842 \\
\textit{Observe your surroundings} \textbf{ (ESE)} & 0.842& 0.947& 0.972& 0.972& 0.842& 0.972& 0.842& 0.895& 0.947& 0.972 \\
\textit{Play mobile game} \textbf{ (CA)} & 0.263*& 0.105*& 0.842& 0.474& 0*& 0.263*& 0.053*& 0.368*& 0.263*& 0.737 \\
\textit{Scroll through old memories from your gallery} \textbf{ (ESE)} & 0.421& 0.316*& 0.947& 0.842& 0.158*& 0.684& 0.158*& 0.526& 0.632& 0.895 \\
\textit{Stretching} \textbf{ (PA)} & 0.368*& 0.947& 0.972& 0.158*& 0.053*& 0.737& 0.368*& 0.895& 0.105*& 0.947 \\
\textit{Watching funny videos} \textbf{ (MR)} & 0.158*& 0.053*& 0.972& 0.842& 0.211*& 0.789& 0*& 0.368*& 0.842& 0.842 \\
\textit{Watching motivational video} \textbf{ (MR)} & 0.158*& 0.579& 0.842& 0.684& 0.211*& 0.579& 0.105*& 0.526& 0.632& 0.789 \\
\textit{Writing Down a Worry and Putting It Aside} \textbf{ (CA)} & 0.632& 0.211*& 0.895& 0.368*& 0*& 0.158*& 0*& 0.789& 0.158*& 0.684\\ \bottomrule
\end{tabular}
\end{table}

\subsection{Evaluation metrics}
We followed the methodology of Kunzler and Mishra et al. \cite{kunzler2019exploring} to evaluate the relationship between intervention notifications and passively sensed smartphone data. Specifically, we used two key metrics: \textbf{completion rate} and \textbf{response time}. The completion rate is calculated as the total number of responded notifications divided by the sum of responded and missed notifications. This metric reflects the participant's likelihood of engaging with the intervention. Response time is defined as the time the participant took to respond to a notification and was calculated as the difference between the response and notification timestamps. Shorter response times generally indicate higher receptivity.
To assess the feasibility of adaptive interventions, we used \textbf{average reward} as the evaluation metric. In our setup, a higher average reward indicates that the intervention was both accepted and feasible in the given context. This reward-based evaluation aligns with reinforcement learning frameworks, where maximizing the average reward helps tailor interventions that are more likely to be effective and contextually appropriate for each user.

\section{Data preparation}
The passively collected smartphone data originated from various onboard sensors and usage patterns, with each type of data collected independently. This section describes the preprocessing steps undertaken for each dataset prior to conducting in-depth analyses. A summary of the extracted features used in our study is provided in Table \ref{tab:summary_of_feature}.


\subsection{Date and Time Context}
From the existing literature, it is known that the time of day can significantly influence the acceptance of interventions in mHealth \cite{chaudhari2022personalization}. Therefore, we tested this on our collected mental health data as well. We divided the notification timing between 7:00 AM and 8:00 PM into three categories: Morning (7:00 AM to 12:00 PM), Afternoon (12:00 PM to 4:00 PM), and Evening (4:00 PM to 8:00 PM). We also examined the relationship across different days of the week to see how each day affects intervention acceptance. The rationale for testing for each day comes from prior studies, where the authors found that both positive and negative moods are associated with the days of the week \cite{stone2012day}.

Further, we categorized the days of the week into two types of subcategories. Subcategory 1 included Early Week (Monday and Tuesday), Mid Week (Wednesday, Thursday, and Friday), and Weekend (Saturday and Sunday). Subcategory 2 grouped the days into Weekdays (Monday to Friday) and weekends (Saturday and Sunday). The rationale for grouping days into these buckets comes from prior studies. Ryan et al. \cite{ryan2010weekends} found that weekends are associated with several indicators of well-being. Helliwell et al. \cite{helliwell2014weekends} also observed that people experience happiness, enjoyment, and laughter, and significantly less anxiety, sadness, and anger on weekends. Moreover, building on these studies, we also tried to understand the effect of early week, mid-week, and weekends.

\subsection{Phone Battery Status}
Prior studies suggest that battery levels influence user behavior and experience with smartphones \cite{tang2020quantifying, rahmati2007understanding}. Mobile operating systems often restrict various application functionalities to conserve battery, and users also adjust their usage patterns to save battery life. With advancements in operating system algorithms, a power-saving mode has been introduced, which users enable any time to limit mobile and app functionalities. 


Our LogMe app collected three important battery-related factors along with their logged timestamps: battery level (0–100\%), battery status (Charging, Full, Discharging, Not Charging), and power saving mode (True/False). The battery level was categorized into four groups based on its value: 'Critical' (0–10\%), 'Low' (10–20\%), 'Medium' (20–60\%), and 'High' (60–100\%). The rationale behind this categorization comes from how mobile operating systems alert users about battery status. Typically, users receive a notification to turn on power-saving mode when the battery reaches around 20\%, indicating a low battery, and around 10\%, indicating a critical battery level.

We mapped the timestamps of intervention notifications to the corresponding battery data. If an exact timestamp match was not available, we mapped it to the nearest battery timestamp. We then removed any mapped data where the time difference between the notification timestamp and the battery log exceeded 30 minutes.

\subsection{Screen Interaction Patterns}
Screen interaction provides insight into how often a participant engages with their phone. We collected two subcategories of screen interaction data: screen status (On/Off) and unlock event (True/False). Screen status indicates whether the screen was on or not, while the unlock event shows whether the user unlocked the phone. Based on these two factors, three possible cases can occur: (i) Screen On and Unlock Event False, where the user presses the power button or lifts the phone to check the time or notifications without unlocking it; (ii) Screen On and Unlock Event True, where the user lifts the phone or presses the power button and unlocks it; and (iii) Screen Off and Unlock Event False, where the screen times out and locks automatically, or the user manually presses the power button to lock the phone. 

Screen interaction data was collected only when there was a change from the previous state. Analyzing screen interaction was important to understand how participants responded to intervention notifications. Similar to the battery mapping process with intervention notifications, we mapped the screen interaction timestamps to the nearest intervention notification timestamps. A tolerance of 30 minutes was allowed for this mapping. Here, tolerance means that mappings were excluded if the time difference between the intervention notification and the nearest screen interaction exceeded 30 minutes.


\subsection{Physical Activity Recognition}
The physical activity of a user often influences their interaction with smartphones. For example, a user might ignore a notification while running, or miss a notification entirely if they are walking with the phone in their pocket. Therefore, analyzing physical activity can help identify the activities during which users are more likely to notice and respond to notifications. Using Google Activity Recognition API, participant activities were categorized as Still, On Foot, Walking, Tilting, In Vehicle, Running, and On Bicycle. The system leverages Google’s on-device machine learning models to classify these activities. Several studies have validated the accuracy of this API in real-world settings. For instance, a large-scale fall detection study using the API reported 73\% sensitivity and >99.9\% specificity across over 2,000 participant-days \cite{harari2021smartphone}. Another study modeling human activity reported a training accuracy of 84\% and validation accuracy of 71\%, confirming its effectiveness for behavioral classification using smartphones \cite{badrinath2021modelling}.

Based on existing literature, we combined `On Foot' and `Walking' into a single category \cite{kunzler2019exploring, mishra2020evaluating}. Each activity provides specific insights into participant behavior; for instance, `Still' indicates that the participant is stationary or the phone is placed somewhere, while `Tilting' suggests that the device is being lifted or actively used. We mapped the timestamps of intervention notifications to the corresponding logged physical activity timestamps. A five-minute window was used for this mapping, where if a notification was sent at time \textit{t+5}, the window considered was from \textit{t0} to \textit{t+5}. Within this window, we identified the mode of all physical activities recorded. The five-minute window size was selected empirically after a series of hit-and-trial experiments with varying window durations.

\subsection{App Usage Behavior} \label{subsection:app_usage_behavior}
User app usage behavior can influence how they respond to intervention notifications. For example, a user engaged in gaming or performing important tasks on their phone may ignore an intervention notification. Therefore, it is important to understand app usage behavior to identify the application usage during which users are more likely to respond promptly to intervention notifications. Our LogMe data collection application recorded each session's (whenever an app is opened) timestamp, app package name, and foreground duration. To identify the application names from the package names, we first checked whether the applications were available on the Google Play Store. Some package names were not found in the Play Store; upon closer inspection, it was found that these packages corresponded to system applications such as Android Launcher, Oplus Wireless Settings, or brand-specific factory applications that are not publicly available on the Play Store. In total, 610 applications were identified. These applications were grouped into seven categories based on their function, with `Other' as one category. The final categories were Other, Productivity \& Tools, Communication \& Social, Games \& Simulation, Entertainment \& Media, Shopping, Finance \& Travel, and Lifestyle \& Health. 


\subsection{Call Activity}
The call status of a user indicates whether they were on a call at the time a notification was dropped. Call details are important because users may reject or delay interaction with notifications if they are engaged in important calls. Our LogMe application captured call-related information, including Call Type (incoming or outgoing), Call Start Time, Call End Time, Call Duration, and Call Status (missed if the user did not answer an incoming call or complete if the call was answered or if a user made the call). We mapped the intervention notification timestamps to the call timestamps. For each participant, an intervention notification was labeled as `Yes' (on call) if the notification timestamp fell between the Call Start and Call End times, indicating that the user was on a call when the notification was delivered. Otherwise, the intervention was labeled as `No' (not on call).

\subsection{Location Context (GPS)} \label{subsection:location}
The location of a user can also help in identifying whether a user is likely to interact with a notification. For example, if a user is in a sports area, they may have kept their phone aside, leading to a missed or delayed response to the notification. Our LogMe application collected location-related data, including Timestamp, Latitude, and Longitude. Initially, we implemented an event-driven GPS data collection method, where a new data point was logged only if there was a change in location compared to the previous entry. However, this approach led to high battery consumption, so we shifted to a fixed sampling rate of one data point every 15 minutes. We used the Google Places API to find a valid location name associated with each collected latitude and longitude. A tolerance of 50 meters was set, prioritizing the closest valid location within this range. We then mapped the intervention notification timestamps with the nearest GPS location timestamps, allowing a maximum tolerance of 15 minutes. A total of 211 valid location names were identified and categorized into six main groups: Campus Open Area, Academic Building or Lab, Sports Region, Cafeteria \& Eatery, Dormitory Area, and Outside Campus.


\begin{table}[!ht]
\centering
\scriptsize
\caption{Summary of extracted features from smartphone data.}
\begin{tabular}{p{2.5cm}p{3cm}p{7cm}p{0.5cm}}
\toprule
\textbf{Mobile Usage} & \textbf{Features} & \textbf{Factors} & \textbf{\#} \\ \midrule
\multirow{4}{*}{Date/Time} 
    & Time of Day & Categorical (Morning, Afternoon, Evening) & 3 \\
    & Type of Day & Categorical (Monday, Tuesday, Wednesday, Thursday, Friday, Saturday, Sunday) & 7 \\
    &  Week type I& Categorical (Early week, Mid week, Weekend)& 3 \\
    &  Week type II& Categorical (Weekdays / Weekend) & 2 \\ \midrule
\multirow{3}{*}{Phone Battery} 
    & Status & Categorical (Discharging, Not Charging, Charging, Full) & 4 \\
    & Level & Categorical (Critical, Low, Medium, High) & 4 \\
    & Power Saving & Categorical binary (True / False) & 2 \\ \midrule
\multirow{2}{*}{Screen Interaction} 
    & Lock State & Categorical binary (True / False) & 2 \\
    & Screen Status & Categorical binary (Off / On) & 2 \\ \midrule
Activity & Recognized activity & Categorical (Still, Tilting, Walking, In Vehicle, On Bicycle, Running) & 6 \\ \midrule
App Usage & App Category & 
    Categorical (Other, Communication \& Social, Games \& Simulation, Productivity \& Tools, Entertainment \& Media, Shopping, 
    Finance \& Travel, Lifestyle \& Health) & 6 \\ \midrule
Calls & On Call & Categorical binary (Yes / No) & 2 \\ \midrule
GPS & Location Category & Categorical (Campus Open Area, Academic Building \& Lab, Sports Region, Cafeteria \& Eatery, Dormitory Area, Outside Campus) & 6 \\
\bottomrule
\end{tabular}%
\label{tab:summary_of_feature}
\end{table}

\section{Data analysis}
To address our research questions, we present the analysis and findings using a combination of data visualizations and inferential statistics. The choice of statistical tests was guided by the nature of the dependent and independent variables under comparison. As the data did not meet the assumptions of normality required for parametric tests, we employed non-parametric alternatives. Specifically, the Kruskal-Wallis test was used for comparisons involving three or more factors, serving as a non-parametric equivalent to ANOVA. For comparisons between the two factors, we applied the Mann-Whitney U test. When post hoc analysis was required, we used the Dunn test with Bonferroni correction for p-value adjustment.
The analysis for RQ1 is presented in Section \ref{section: analysis_1}, while RQ2 and RQ4 are detailed in Section \ref{section: analysis_2}. Analysis corresponding to RQ3 and additional insights for RQ4 are also included in Section \ref{section: analysis_3}.

\begin{table}[]
\centering
\caption{Summary of filled and missed interventions after mapping intervention notifications to sensed data.}
\label{tab:counts_filled_missed_intervention}
\begin{tabular}{@{}lcc@{}}
\toprule
\textbf{} & \textbf{Filled  
 (\#)} & \textbf{Missed (\#)} \\ \midrule
\textit{Total number of intervention notifications} & 8162 & 2525 \\
\textit{Battery} & 8156 & 2514 \\
\textit{Screen interaction} & 7946 & 2103 \\
\textit{Google activity recognition} & 8161 & 2525 \\
\textit{App Usage} & 7482 & 2070 \\
\textit{Call} & 8162 & 2525 \\
\textit{GPS} & 4744 & 1591 \\ \bottomrule
\end{tabular}
\end{table}



\subsection{Effect of passive data on the Receptivity of intervention notification (RQ1)} \label{section: analysis_1}

The table \ref{tab:counts_filled_missed_intervention} shows the number of intervention notifications that were either responded to (`\textit{Filled}') or not (`\textit{Missed}'), along with the availability of corresponding passive sensing modalities used for analysis in the following subsection.

\subsubsection{Date and Time Context}
We observed that the time of day had a significant effect on the completion rate of intervention notifications, H(2) = 22.127, p < .001. This suggests that there was a significant difference in completion rates across different times of the day. A post hoc test revealed that the completion rate in the morning significantly differed from both the afternoon (p < .001) and evening (p < .001). Upon closer inspection, we found that the completion rate was lower in the morning (7:00 AM to 12:00 PM) compared to the afternoon (12:00 PM to 4:00 PM) and evening (4:00 PM to 8:00 PM) periods (see Figure \ref{fig:CR_DT}). However, no significant difference was found between the afternoon and evening completion rates (see Figure \ref{fig:CR_DT}). Similarly, we observed that the time of day had a significant effect on the response time for intervention notifications, H(2) = 47.698, p < .001. The post hoc test revealed that the morning period significantly differed from both the afternoon (p < .001) and evening (p < .001) in terms of response time. Specifically, we found that response times were higher in the morning compared to the afternoon and evening (see Figure \ref{fig:RR_DT}).

In the case of the type of day (Monday, Tuesday, ..., Saturday, Sunday), we did not find any significant difference in the completion rate, H(6)=12.482, p = 0.052. Similarly, no significant difference was observed for response delay across different days, H(6)=5.344, p = 0.5. This suggests that the type of day does not affect the response rate (see Figure \ref{fig:CR_DW}) or the response delay (see Figure \ref{fig:RR_DW}) of the sent intervention notifications.

When grouping days into three subcategories, i.e., early week, midweek, and weekend, we found a significant difference in completion rate, H(2) = 8.746, p = 0.014. Post hoc analysis revealed that the weekend significantly differed from the early week (p = 0.023) and midweek (p = 0.042), with no significant difference between the early week and midweek. Upon further examination (see Figure \ref{fig:CR_WWI}), we observed that completion was higher during the early week and lower during the weekend. However, no significant difference was found among the subcategories regarding response time. However, Figure \ref{fig:RR_WWI} shows that response time was lower during the early week and higher during the midweek.

Similarly, when grouping the days into weekdays and weekends, we observed a significant difference in completion rate, p=0.004, with lower completion during the weekend than on weekdays (see Figure \ref{fig:CR_WWII}). However, there was no significant difference in response time between weekdays and weekends. Figures \ref{fig:CR_WWII} and \ref{fig:RR_WWII} illustrate that the completion rate was lower and the response time was higher on weekends than on weekdays.


\subsubsection{Phone Battery Status}
\begin{figure}[h]
    \centering
    \begin{subfigure}{0.3\textwidth}
        \includegraphics[width=\linewidth]{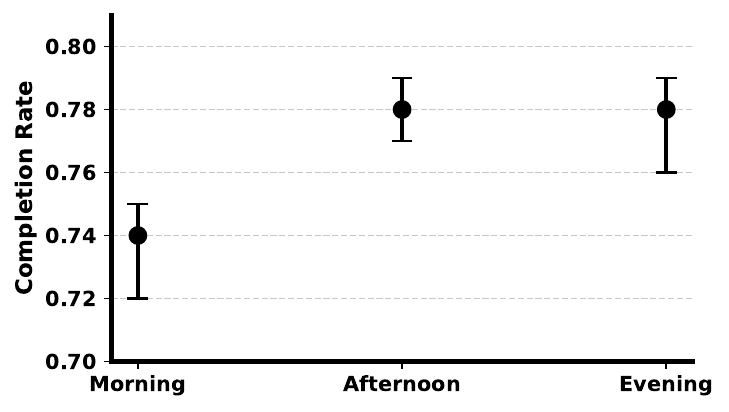}
        \Description{}
        \caption{Timing of Day}
        \label{fig:CR_DT}
    \end{subfigure}
    \hfill
    \begin{subfigure}{0.3\textwidth}
        \includegraphics[width=\linewidth]{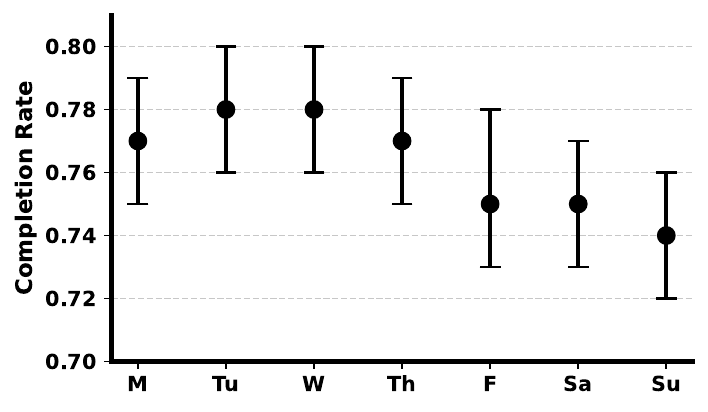}
        \Description{}
        \caption{Day wise}
        \label{fig:CR_DW}
    \end{subfigure}
    \hfill
    \begin{subfigure}{0.3\textwidth}
        \includegraphics[width=\linewidth]{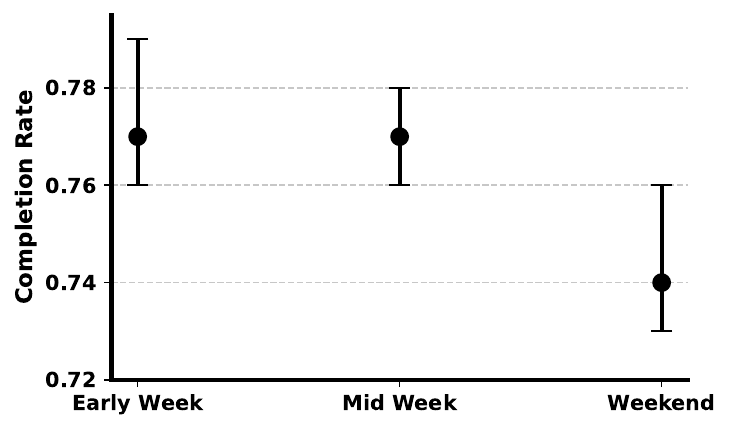}
        \Description{}
        \caption{Week Category I}
        \label{fig:CR_WWI}
    \end{subfigure}
    \begin{subfigure}{0.3\textwidth}
        \includegraphics[width=\linewidth]{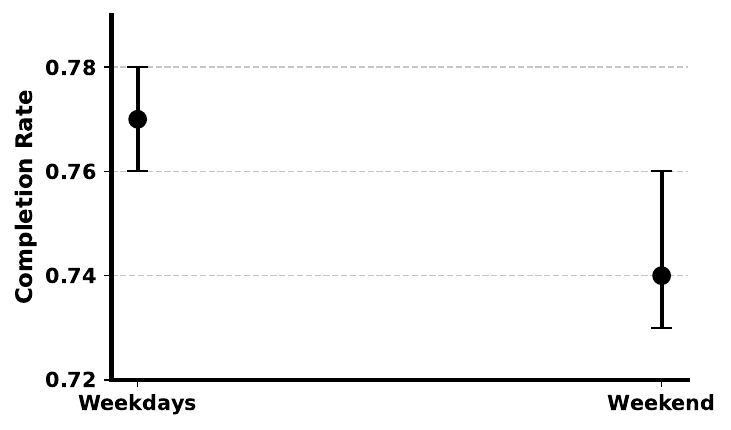}
        \Description{}
        \caption{Week Category II}
        \label{fig:CR_WWII}
    \end{subfigure}
    \hfill
    \begin{subfigure}{0.3\textwidth}
        \includegraphics[width=\linewidth]{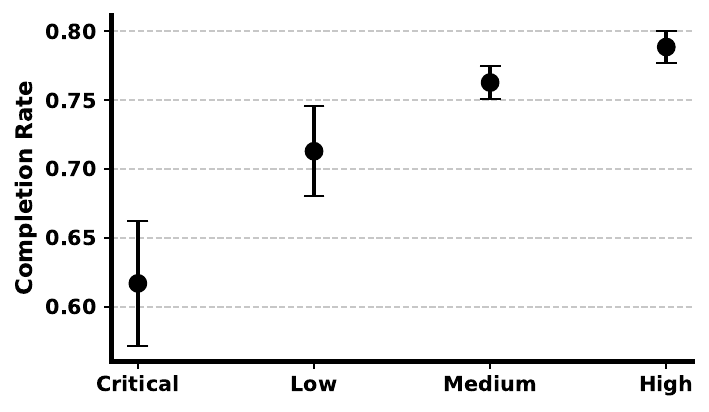}
        \Description{}
        \caption{Battery Level}
        \label{fig:CR_BL}
    \end{subfigure}
    \hfill
    \begin{subfigure}{0.3\textwidth}
        \includegraphics[width=\linewidth]{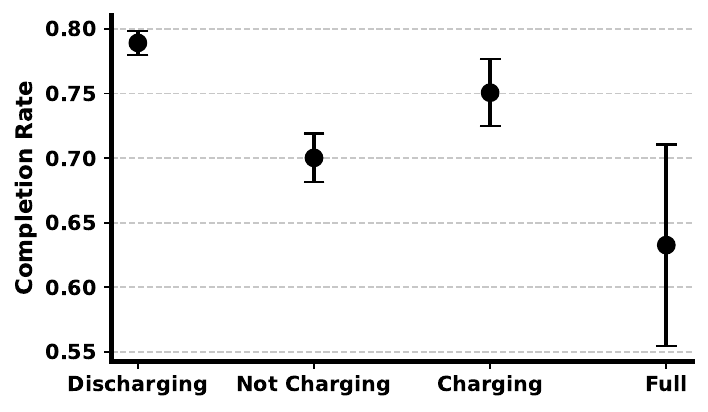}
        \Description{}
        \caption{Battery Status}
        \label{fig:CR_BS}
    \end{subfigure}
    \begin{subfigure}{0.3\textwidth}
        \includegraphics[width=\linewidth]{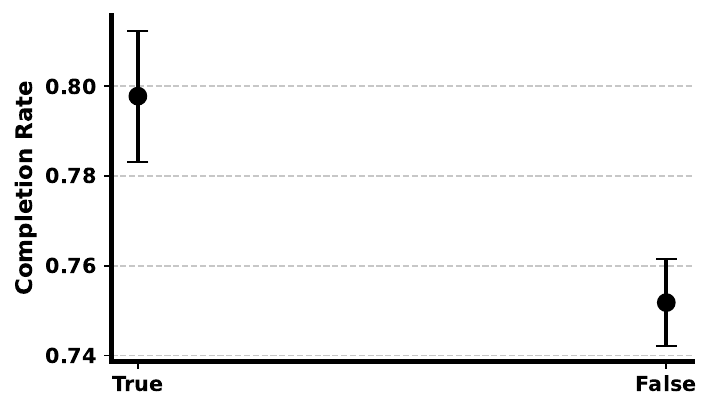}
        \Description{}
        \caption{Battery Power saving mode}
        \label{fig:CR_BPSM}
    \end{subfigure}
    \hfill
    \begin{subfigure}{0.3\textwidth}
        \includegraphics[width=\linewidth]{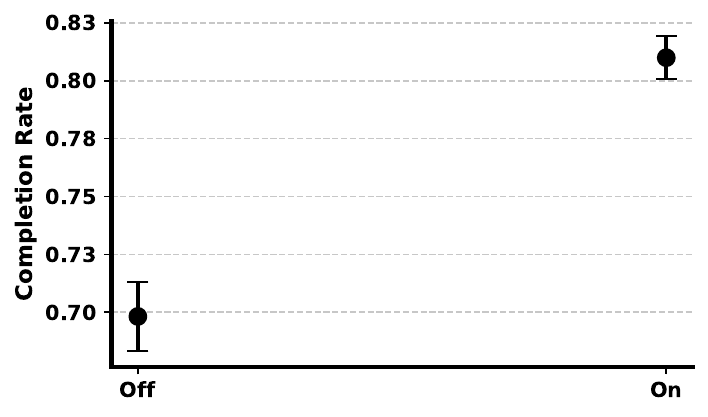}
        \Description{}
        \caption{Device Screen Status}
        \label{fig:CR_SS}
    \end{subfigure}
    \hfill
    \begin{subfigure}{0.3\textwidth}
        \includegraphics[width=\linewidth]{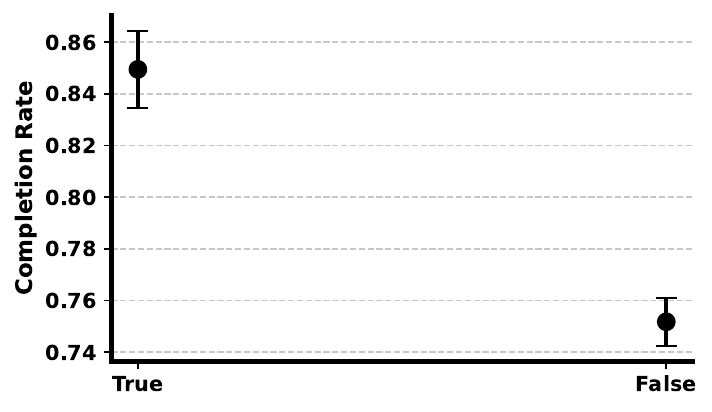}
        \Description{}
        \caption{Device Lock/Unlock event}
        \label{fig:CR_SUE}
    \end{subfigure}
    \begin{subfigure}{0.3\textwidth}
        \includegraphics[width=\linewidth]{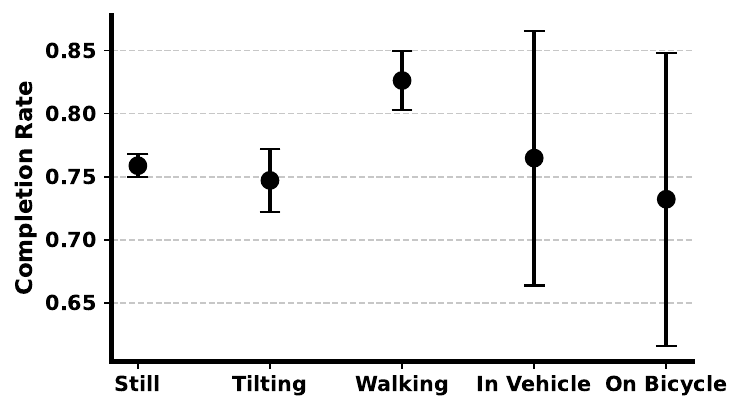}
        \Description{}
        \caption{Activity}
        \label{fig:CR_Gapi}
    \end{subfigure}
    \hfill
    \begin{subfigure}{0.3\textwidth}
        \includegraphics[width=\linewidth]{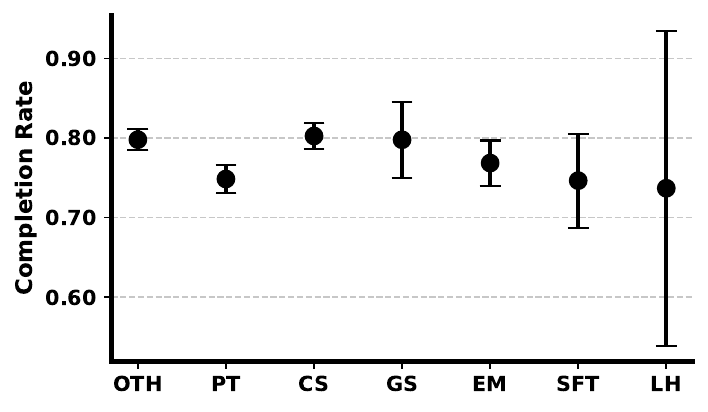}
        \Description{}
        \caption{App usage (category)}
        \label{fig:CR_AU}
    \end{subfigure}
    \hfill
    \begin{subfigure}{0.3\textwidth}
        \includegraphics[width=\linewidth]{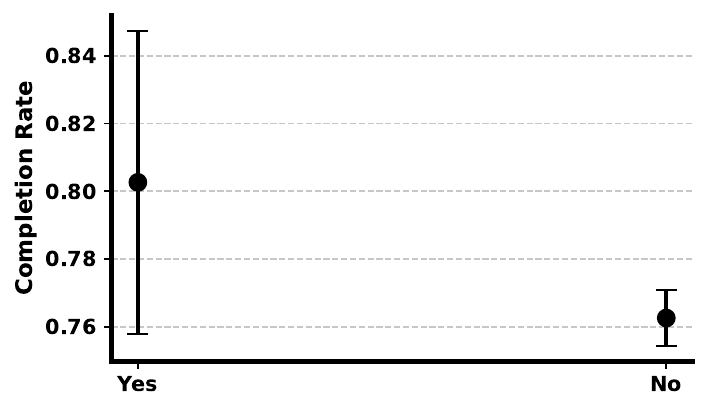}
        \Description{}
        \caption{OnCall}
        \label{fig:CR_Calls}
    \end{subfigure}
    \begin{subfigure}{0.3\textwidth}
        \includegraphics[width=\linewidth]{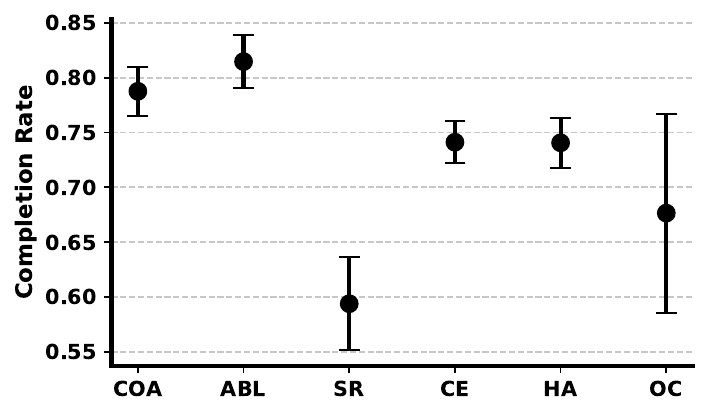}
        \Description{}
        \caption{Location (GPS)}
        \label{fig:CR_gps}
    \end{subfigure}
    \caption{Visual analysis of intervention completion rates under various contextual factors. The abbreviations COA: Campus Open Area, ABL: Academic Building \& Labs, SR: Sports Region, CE: Cafeteria \& Eatery, HA: Hostel Area, OC: Outside Campus, OTH: Other, PT: Productivity \& Tools, CS: Communication \& Social, GS: Games \& Simulation, EM: Entertainment \& Media, SFT: Shopping, Finance \& Travel, LH: Lifestyle \& Health}
    \label{fig:main}
\end{figure}

Next, we examined the relationship between device battery level and receptivity. Battery level showed a significant effect on the acceptance of notifications, measured by the completion rate (H(3) = 79.228, p < .001). Post hoc analysis revealed that the `Critical' battery category significantly differed from the `Low' (p<0.001), `Medium' (p < .001), and `High' (p < .001) categories. Similarly, `Low' significantly differed from `Medium' (p = 0.018), and `High' (p < .001), and `Medium' significantly differed from `High' (p < .018). Figure (\ref{fig:CR_BL}) suggests that higher battery levels are associated with higher completion rates.
Device battery level also showed a significant effect on response time (H(3) = 31.181, p < .001). Post hoc results indicated that response times for 'Critical' and `Low', as well as `Medium' and `High', did not differ significantly from each other. However, `Medium' only significantly differed from critical (p < .001), while `High' differed significantly from `Critical' (p < .001) and `Low' (p = 0.008). Figure (\ref{fig:RR_BL}) shows that although there was a visible difference in response time between `Critical' and `Low' battery levels, the difference was not statistically significant. Furthermore, the response times at `Medium' and `High' battery levels were approximately the same.

We also found a significant effect of phone battery status on both completion rate (H(3) = 91.799, p < .001) and response time (H(3) = 46.209, p < .001). Post hoc analysis revealed that the `Charging' status significantly differed from `Discharging' (p=0.034, 0.001), `Full' (p = 0.009, < .001), and `Not Charging' (p = 0.008, 0.013) statuses for both completion rate and response time, respectively. Additionally, `Discharging' significantly differed from `Full' (p < .001) and `Not Charging' (p < .001) in terms of completion rate, but differed only from `Full' (p < .001) in terms of response time. The `Full' status significantly differed from `Not Charging' (p < .001) in response time but not in completion rate. From Figure (\ref{fig:CR_BS}), we observed that the completion rate was highest when the phone was in the `Discharging' state and lowest when it was `Full.' An interesting pattern was found for response time, where response time (see Figure \ref{fig:RR_BS}) was highest when the battery was `Full' and lowest when the battery status was `Not Charging'.

We found a significant effect of power-saving mode on the completion rate (p < .001), but not on response time. Figure (\ref{fig:CR_BPSM}) shows the completion rate (approx. 80\%) was higher when the phone’s power saving mode was ON. Although Figure (\ref{fig:RR_BPSM}) indicates that response time was higher when the power saving mode was ON than when it was OFF, the difference in response time was not statistically significant.



\subsubsection{Screen Interaction Patterns}
Further, we investigated how different screen interactions affect the acceptance of intervention notifications. We observed a significant effect of both screen lock state (p < .001) and screen status (p < .001) on completion rate. Similarly, we found a significant effect of screen lock state (p < .001) and screen status (p < .001) on response time. Figure  (\ref{fig:CR_SS}) shows that when the screen status was ON, the completion rate was higher than when the screen was OFF, with a completion rate of approximately 82\%. Likewise, when the phone was unlocked, the completion rate was approximately 85\%, higher than when the phone was locked. A similar pattern was observed for response time, where the response time was lower when the screen was ON and the phone was unlocked.

\subsubsection{Physical Activity Recognition}
Next, we evaluated the effect of participants' activity, detected using Google Activity Recognition, on the intervention notification. We observed a significant effect of detected activity on both completion rate (H(5) = 25.287, p < .001) and response time (H(5) = 27.938, p < .001). In the post hoc test, we found that only the activities Still (p < .001) and Tilting (p < .001) significantly differed from Walking for both completion rate and response time. Figure (\ref{fig:CR_Gapi}) shows that the completion rate during Walking was higher, approximately 83\%. Similarly, Figure (\ref{fig:RR_Gapi}) shows that the response time was lowest during Walking.


\subsubsection{App Usage Behavior}
Next, we investigated the effect of app usage behavior on the sent intervention notifications. We grouped the apps used by the participants into six main categories, as discussed in Section \ref{subsection:app_usage_behavior}. We found a significant effect of app category usage on completion rate (H(5) = 29.722, p < .001). Further, the post hoc test revealed that Communication \& Social applications significantly differed from Productivity \& Tools applications (p < .001). Additionally, the Other category significantly differed only from Productivity \& Tools applications (p < .001). Figure (\ref{fig:CR_AU}) shows that the lowest completion rate was observed when users were using Communication \& Social applications.

Similarly, we also found a significant effect of app category usage on response time (H(5) = 163.099, p < .001). The post hoc test suggested that Communication \& Social applications significantly differed from Entertainment \& Media (p < .001), Games \& Simulation (p < .001), Other (p < .001), and Productivity \& Tools (p < .001) categories. Entertainment \& Media applications significantly differed from Productivity \& Tools applications (p < .001). Additionally, the Other category differed significantly from Productivity \& Tools (p < .001). Figure (\ref{fig:RR_AU}) shows that the lowest response time was observed while users were using Communication \& Social apps, while the highest response time was observed during Games \& Simulation app usage.


\subsubsection{Call Activity}
Calls are an important factor, as users are already engaged with their mobile devices and can notice haptic feedback (i.e., intervention notification) more easily. In this analysis, we observed that the On Call status (i.e., Yes/No) did not significantly affect the completion rate. However, Figure (\ref{fig:CR_Calls}) shows that the completion rate appeared higher during On Call periods. Regarding response time, we found a significant effect of On Call status. Specifically, the response time was lower when the user was On Call (Yes) than when they were Not On Call (No) (see Figure \ref{fig:RR_Calls}).


\subsubsection{Location Context (GPS)}
\begin{figure}[h]
    \centering
    \begin{subfigure}{0.3\textwidth}
        \includegraphics[width=\linewidth]{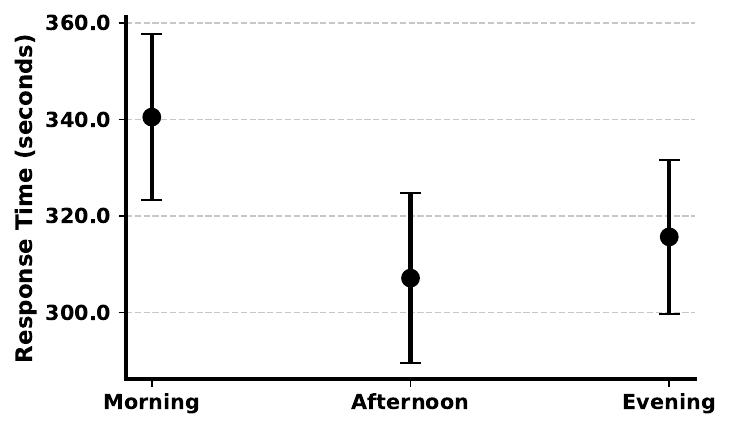}
        \caption{Timing of Day}
        \label{fig:RR_DT}
    \end{subfigure}
    \hfill
    \begin{subfigure}{0.3\textwidth}
        \includegraphics[width=\linewidth]{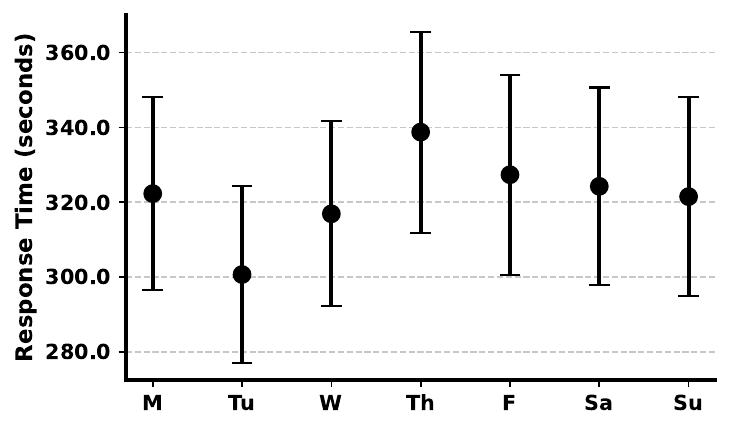}
        \caption{Day wise}
        \label{fig:RR_DW}
    \end{subfigure}
    \hfill
    \begin{subfigure}{0.3\textwidth}
        \includegraphics[width=\linewidth]{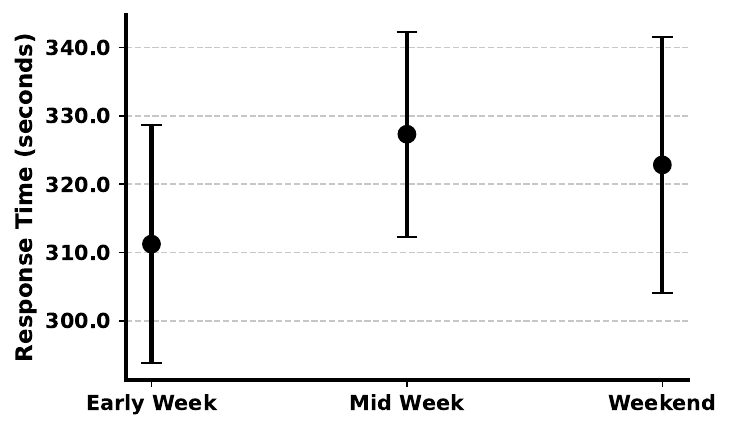}
        \caption{Week Category I}
        \label{fig:RR_WWI}
    \end{subfigure}
    \begin{subfigure}{0.3\textwidth}
        \includegraphics[width=\linewidth]{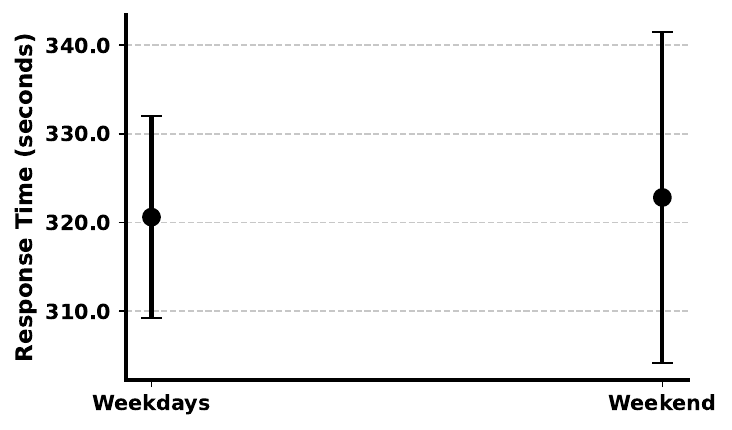}
        \caption{Week Category II}
        \label{fig:RR_WWII}
    \end{subfigure}
    \hfill
    \begin{subfigure}{0.3\textwidth}
        \includegraphics[width=\linewidth]{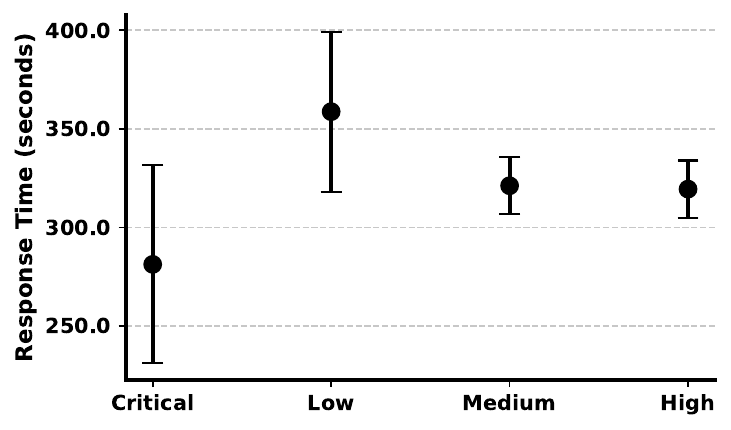}
        \caption{Battery Level}
        \label{fig:RR_BL}
    \end{subfigure}
    \hfill
    \begin{subfigure}{0.3\textwidth}
        \includegraphics[width=\linewidth]{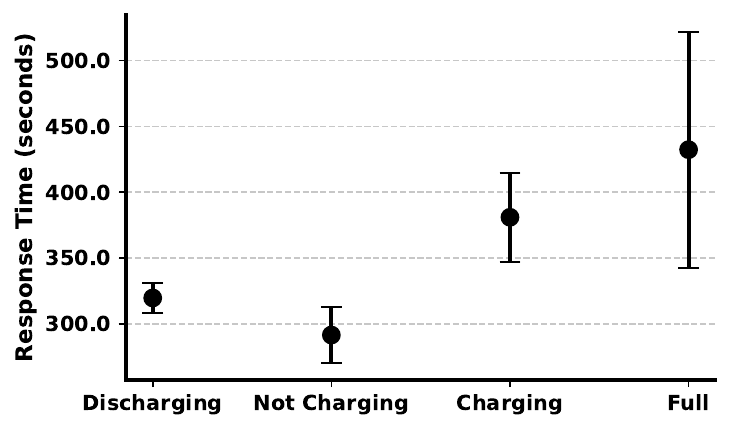}
        \caption{Battery Status}
        \label{fig:RR_BS}
    \end{subfigure}
    \begin{subfigure}{0.3\textwidth}
        \includegraphics[width=\linewidth]{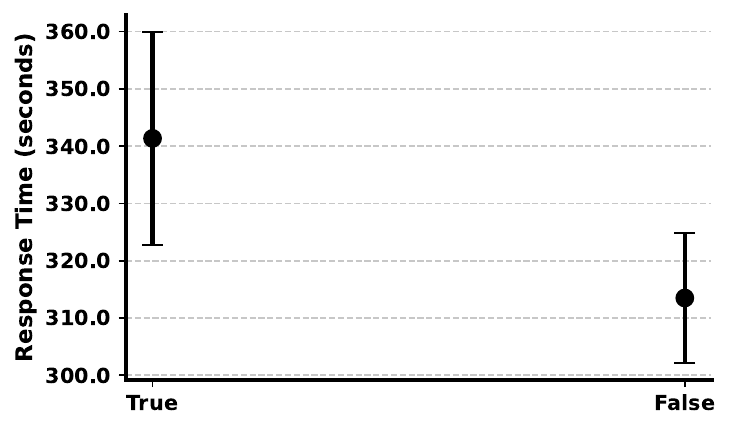}
        \caption{Battery Power saving mode}
        \label{fig:RR_BPSM}
    \end{subfigure}
    \hfill
    \begin{subfigure}{0.3\textwidth}
        \includegraphics[width=\linewidth]{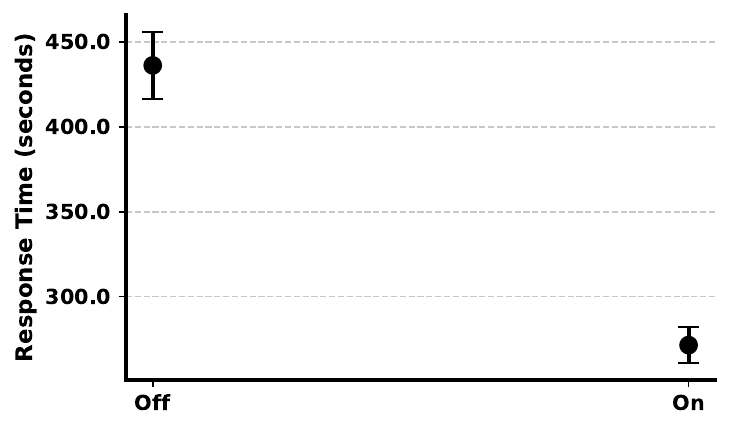}
        \caption{Device Screen Status}
        \label{fig:RR_SS}
    \end{subfigure}
    \hfill
    \begin{subfigure}{0.3\textwidth}
        \includegraphics[width=\linewidth]{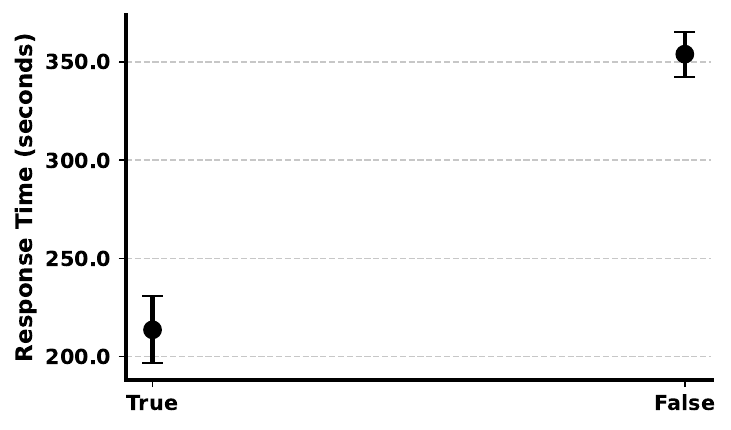}
        \caption{Device Screen Status}
        \label{fig:RR_SUE}
    \end{subfigure}
    \begin{subfigure}{0.3\textwidth}
        \includegraphics[width=\linewidth]{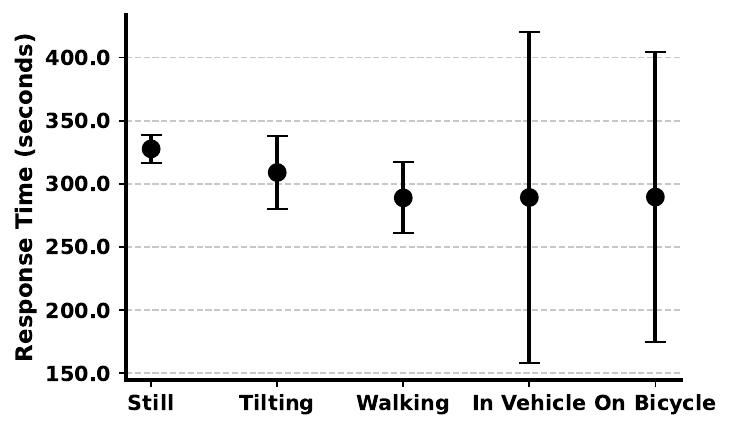}
        \caption{Activity}
        \label{fig:RR_Gapi}
    \end{subfigure}
    \hfill
    \begin{subfigure}{0.3\textwidth}
        \includegraphics[width=\linewidth]{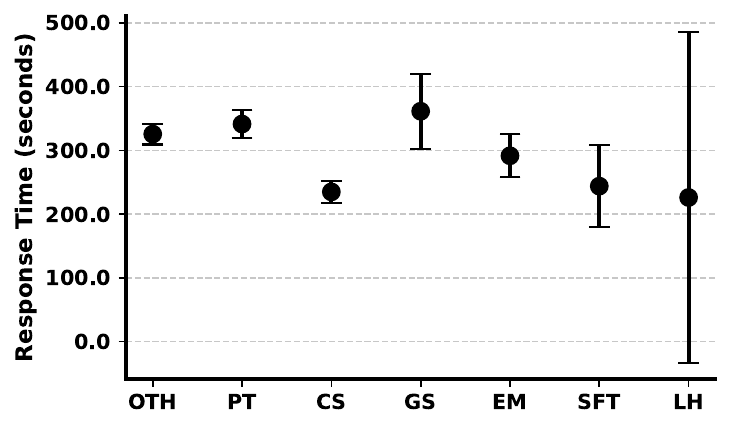}
        \caption{App usage (category)}
        \label{fig:RR_AU}
    \end{subfigure}
    \hfill
    \begin{subfigure}{0.3\textwidth}
        \includegraphics[width=\linewidth]{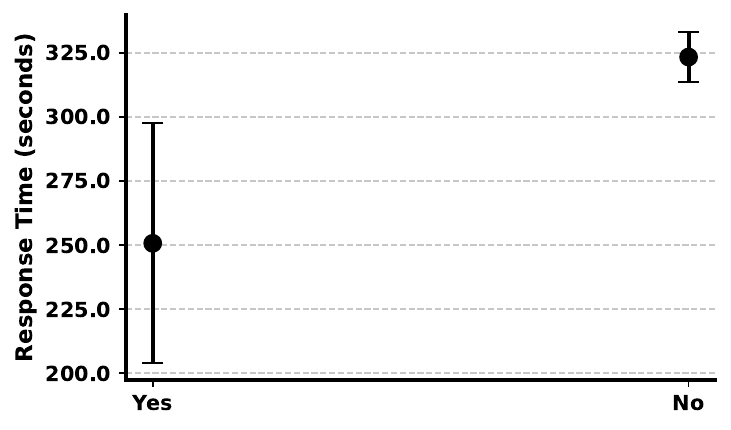}
        \caption{OnCall}
        \label{fig:RR_Calls}
    \end{subfigure}
    \begin{subfigure}{0.3\textwidth}
        \includegraphics[width=\linewidth]{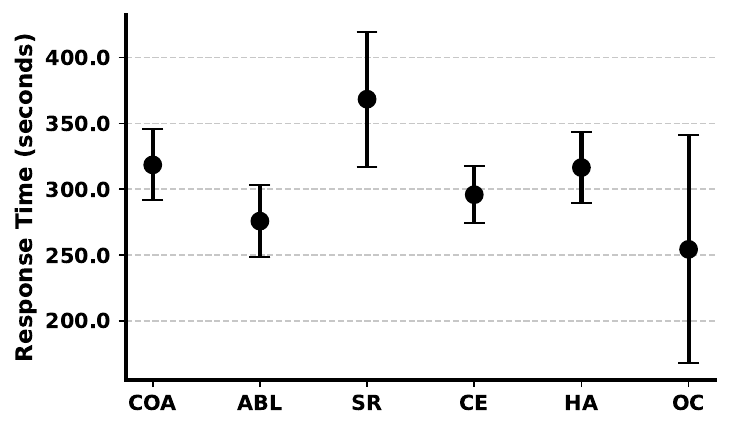}
        \Description{}
        \caption{Location (GPS)}
        \label{fig:RR_gps}
    \end{subfigure}
    \caption{Visual analysis of intervention response times (seconds) under various contextual factors. The abbreviations COA: Campus Open Area, ABL: Academic Building \& Labs, SR: Sports Region, CE: Cafeteria \& Eatery, HA: Hostel Area, OC: Outside Campus, OTH: Other, PT: Productivity \& Tools, CS: Communication \& Social, GS: Games \& Simulation, EM: Entertainment \& Media, SFT: Shopping, Finance \& Travel, LH: Lifestyle \& Health}
    \label{fig:main}
\end{figure}

In our next evaluation, we focused on the effect of participants' locations (extracted from GPS coordinates) on intervention notifications. We grouped the participants' locations into six major categories, as discussed in Section \ref{subsection:location}. Our statistical analysis revealed a significant effect of location on the completion rate, H(5) = 103.146, p < .001. Through post hoc analysis, we found that Academic Buildings/Labs significantly differed from Cafeteria \& Eatery (p < .001), Hostel/dormitory Area (p < .001), Sports Region (p = 0.03), and Outside Campus (p < .001). Cafeteria \& Eatery significantly differed from Campus Open Area (p = 0.04) and Sports Region (p < .001). The Campus Open Area significantly differed from the Sports Region (p < .001), and the Dormitory Area significantly differed from the Sports Region (p < .001). Figure (\ref{fig:CR_gps}) shows that the highest completion rate was observed in Academic Buildings/Labs, while the lowest was in the Sports Region.
We also found a significant effect of location on response time, H(5) = 87.434, p < .001. In the post hoc test, we observed that the Sports Region significantly differed from Academic Buildings/Labs (p < .001), Campus Open Area (p < .001), Cafeteria \& Eatery (p < .001), and Dormitory Area (p < .001). Additionally, a significant difference was found between the Dormitory Area and Academic Buildings/Labs (p < .001). From Figure (\ref{fig:RR_gps}), we observed that although the lowest response time was for Outside Campus locations, it did not significantly differ from other categories. Excluding Outside Campus, the lowest response time was found in Academic Buildings/Labs, while the highest response time was observed in the Sports Region.


\subsection{Effect of passively data on Average Reward: Suggesting feasibility (RQ2 \& RQ4)}  \label{section: analysis_2}
Before deploying any JITAI in real-world settings, it is essential first to understand the feasibility of interventions across different contexts. However, it is important to note that an intervention accepted in one context may not necessarily be accepted again in the same context. For instance, a student experiencing exam-related stress during one exam might accept a breathing exercise to calm down. However, during another exam, even in a similarly stressful situation, the same student might feel too anxious or time-constrained and ignore the suggestion. Similarly, a participant shared at the end of the study that while they would accept a neck roll exercise to relieve academic stress during long study hours when studying alone, they would likely refuse it when studying in a group or in a library due to embarrassment. These examples highlight the need for adaptive interventions rather than fixed interventions.

Moreover, it also remains important to understand the feasibility of interventions based on users' contexts to support the deployment of JITAI models in real-world settings. In this study, participants' responses of ``Yes'' and ``No'' reflect the willingness of the intervention, while ``Yes, but not feasible right now'' provide insight into the participants' personal willingness, constrained by their current environment. Another important application of these findings is to identify the timing when interventions are most acceptable to users, which is the primary objective of this work.


\subsubsection{Date and Time Context}
We observed that daytime significantly affected the average reward, H(2) = 7.720, p = 0.02. The post hoc analysis suggests that the average reward in the evening significantly differed from the morning. Figure \ref{fig:AR_DT} shows that the average reward for the acceptance of interventions was highest in the evening and lowest in the morning. This may be due to mood fluctuations where one feels best in the morning and lower as the day progresses. This could explain the higher acceptance of interventions in the evening, resulting in a higher average reward. A similar finding was reported by Bu et al. \cite{bu2025will}, where they found that university students exhibited better mental health and well-being in the morning.

For individual days of the week, we did not find a significant effect on the average reward. Figure \ref{fig:AR_DW} shows the average reward for each day, where no clear pattern is visible. Similarly, Bu et al. \cite{bu2025will} found little evidence that physiological processes differ significantly across days of the week.

When categorizing days into broader groups, we found no significant effect of either the Week I or Week II categories on the average reward. Although Figures \ref{fig:AR_WWII} and \ref{fig:AR_WWI} show that the average reward tended to be lower on weekends compared to early week, midweek, and weekdays overall, these differences were not statistically significant. This suggests that during weekends, people generally experience a higher positive affect and tend to focus more on their personal time, potentially leading to lower acceptance of interventions \cite{dzogang2017circadian}.

\subsubsection{Phone Battery Status}
In our evaluation of battery-related factors, we found that battery level significantly affected the average reward, H(3) = 8.993, p < .001. However, the post hoc analysis revealed no significant differences when comparing individual battery level categories. Similarly, battery status showed a significant effect on the average reward, H(3) = 78.975, p < .001. The post hoc test for battery status indicated that the average reward for the `Not Charging' state significantly differed from both the `Charging' (p = 0.007) and `Discharging' (p < .001) states. Figure \ref{fig:AR_BS} shows that the highest average reward was observed when the battery status was `Not Charging'. The `Not Charging' status indicates that the phone receives very low power output, such as when connected to a PC or laptop via a USB cable. It is possible that participants were using their PCs or laptops in a seated position while their phones were connected through a USB cable. This scenario might explain why response times were lower in the `Not Charging' state and why there was a higher acceptance of the feasibility of interventions during this condition. Moreover, we found that the battery power-saving mode had a significant effect on the average reward (p = 0.004), suggesting that the feasibility of interventions was higher when the power-saving mode was turned off (see Figure \ref{fig:AR_BPSM}).

\subsubsection{Screen Interaction Patterns}
In the case of screen interaction, we found that screen status did not significantly affect the average reward, although Figure \ref{fig:AR_SS} shows that the average reward was higher when the screen status was OFF. This suggests that the feasibility of interventions is not influenced by screen status. However, we found a significant effect of the screen unlock/lock event on the average reward (p < .001), where the average reward was higher when the unlock event was True (see Figure \ref{fig:AR_SUE}). This indicates that when the phone is unlocked, participants might be more likely to be at leisure and use the suggested interventions, leading to a higher average reward and greater feasibility of interventions.

\subsubsection{Physical Activity Recognition}
Upon evaluating the effect of activity (extracted from Google Activity Recognition) on the feasibility of interventions, we did not find any significant effect. However, data visualization (see Figure \ref{fig:AR_Gapi}) revealed that the feasibility of interventions was higher when the phone was in a tilt state or when the user was walking, compared to when the user was still or in a vehicle. Nevertheless, these differences were not significant enough to be accepted.

\subsubsection{App Usage Behavior}
Next, we evaluated the feasibility of interventions based on the usage of different categories of phone applications. We found that app usage significantly affected the average reward (H(5) = 54.985, p < .001), suggesting that the feasibility of interventions varied across different types of applications. The post hoc analysis indicated that the 'Other' app category significantly differed from the Communication \& Social (p = 0.002) and Entertainment \& Media (p = 0.003). However, since the 'Other' category consisted of a mixture of applications, we could not draw specific conclusions about which app usages led to higher intervention acceptance and, thus, higher feasibility of interventions.
Upon visualizing the results in Figure \ref{fig:AR_AU}, we observed that the highest average reward was in the Lifestyle \& Health category, followed by Games \& Simulation, while the lowest was found in the 'Other' category. It is possible that when participants were using Lifestyle \& Health applications, they were already motivated toward self-care, making them more likely to accept interventions, leading to a higher average reward. However, these differences were not significant enough to be firmly accepted.

\subsubsection{Call Activity}
Next, we evaluated the effect of calls on the feasibility of interventions. The results were not significant enough to be accepted. As shown in Figure \ref{fig:AR_Calls}, the average reward for On Call (Yes) and Not On Call (No) conditions was approximately the same. Although generally a user on a call might not be able to perform an intervention, our results suggest that being on a call or not does not make a noticeable difference in the feasibility of interventions.

\subsubsection{Location Context (GPS)}
\begin{figure}[]
    \centering
    \begin{subfigure}{0.3\textwidth}
        \includegraphics[width=\linewidth]{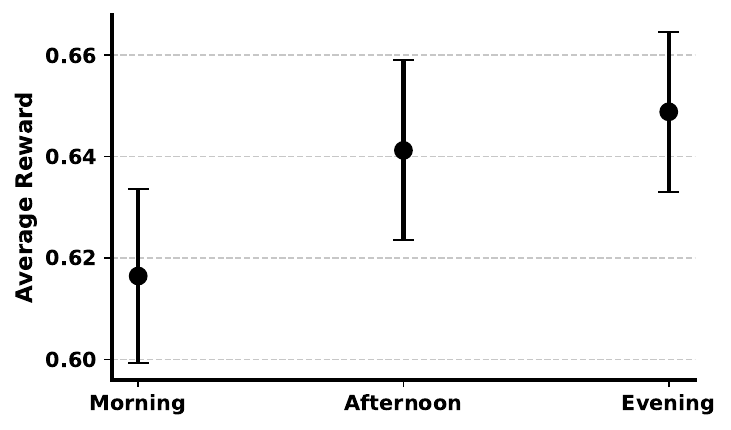}
        \caption{Timing of Day}
        \label{fig:AR_DT}
    \end{subfigure}
    \hfill
    \begin{subfigure}{0.3\textwidth}
        \includegraphics[width=\linewidth]{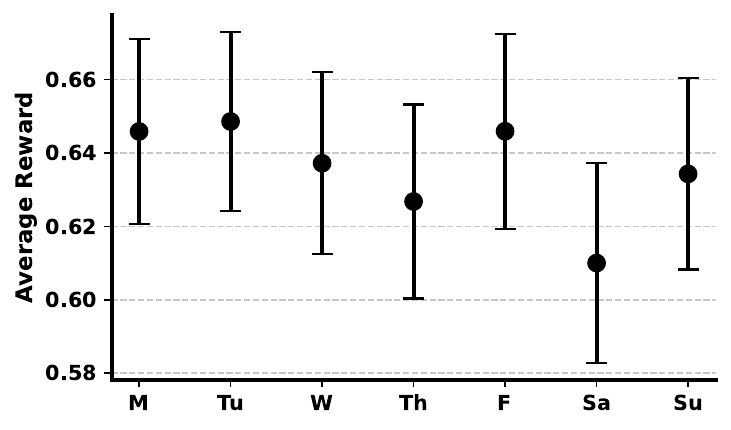}
        \caption{Day wise}
        \label{fig:AR_DW}
    \end{subfigure}
    \hfill
    \begin{subfigure}{0.3\textwidth}
        \includegraphics[width=\linewidth]{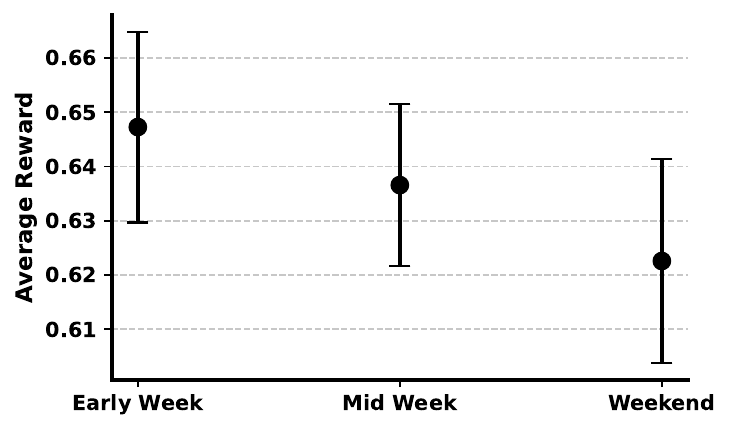}
        \caption{Week Category I}
        \label{fig:AR_WWI}
    \end{subfigure}
    \vspace{0.5cm}
    \begin{subfigure}{0.3\textwidth}
        \includegraphics[width=\linewidth]{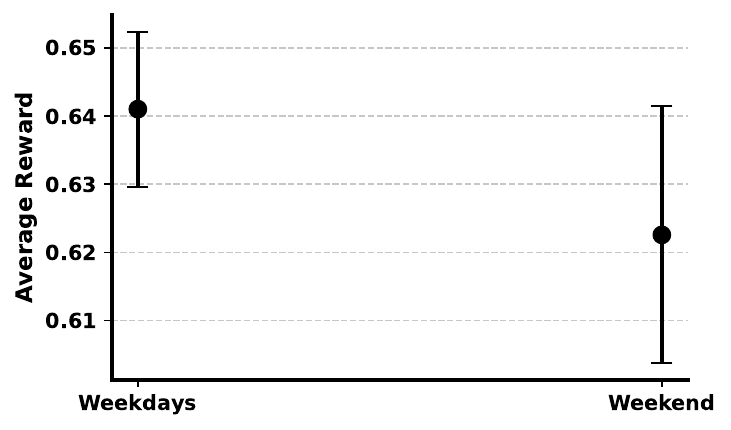}
        \caption{Week Category II}
        \label{fig:AR_WWII}
    \end{subfigure}
    \hfill
    \begin{subfigure}{0.3\textwidth}
        \includegraphics[width=\linewidth]{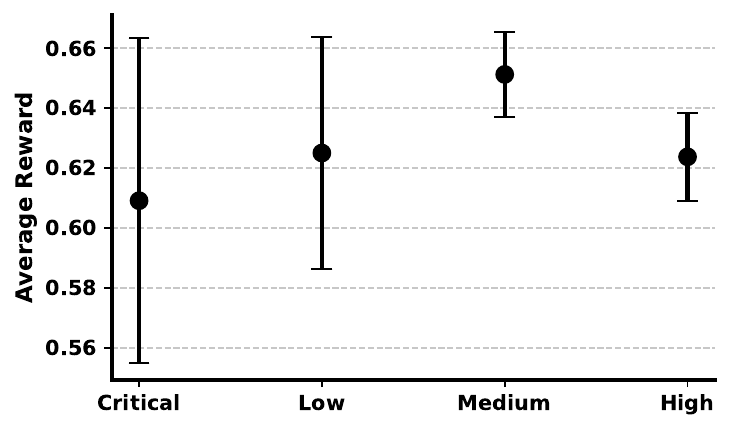}
        \caption{Battery Level}
        \label{fig:AR_BL}
    \end{subfigure}
    \hfill
    \begin{subfigure}{0.3\textwidth}
        \includegraphics[width=\linewidth]{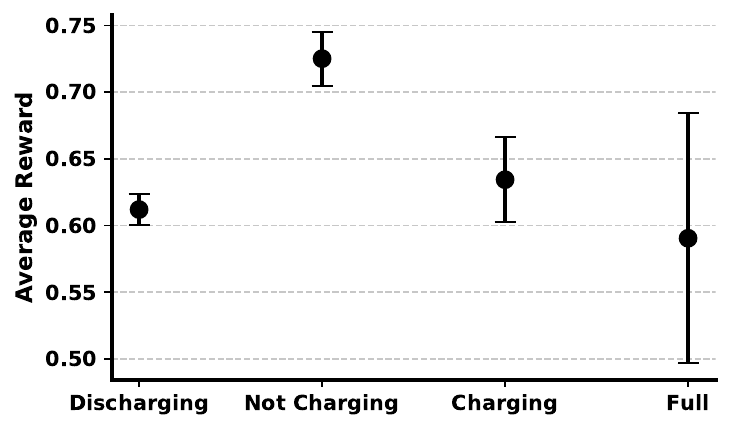}
        \caption{Battery Status}
        \label{fig:AR_BS}
    \end{subfigure}
    \vspace{0.5cm}
    \begin{subfigure}{0.3\textwidth}
        \includegraphics[width=\linewidth]{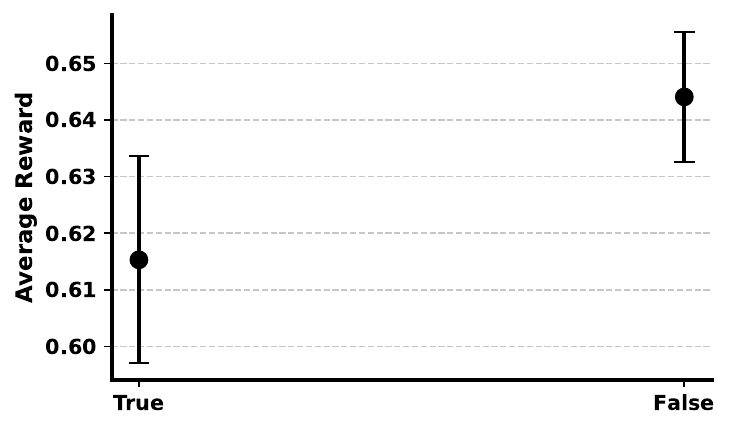}
        \caption{Battery Power saving mode}
        \label{fig:AR_BPSM}
    \end{subfigure}
    \hfill
    \begin{subfigure}{0.3\textwidth}
        \includegraphics[width=\linewidth]{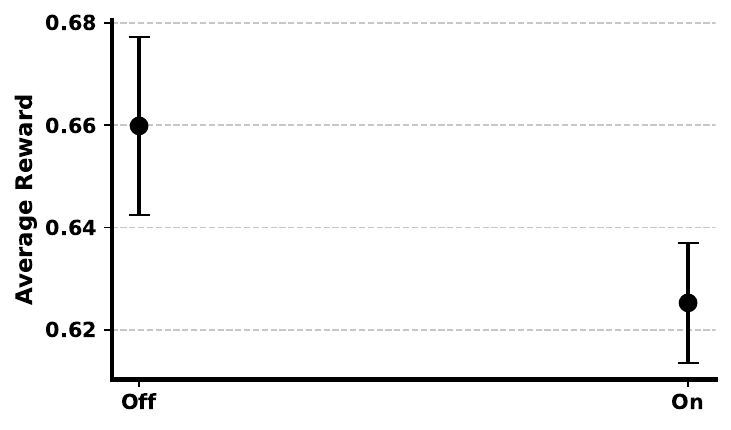}
        \caption{Device Screen Status}
        \label{fig:AR_SS}
    \end{subfigure}
    \hfill
    \begin{subfigure}{0.3\textwidth}
        \includegraphics[width=\linewidth]{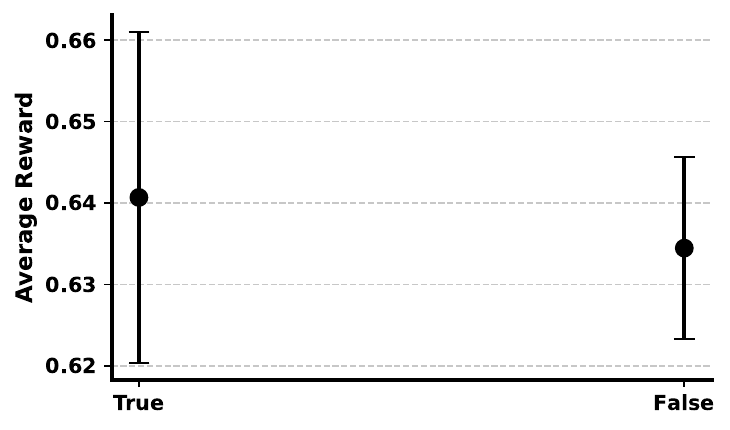}
        \caption{Device Lock/Unlock event}
        \label{fig:AR_SUE}
    \end{subfigure}
    \vspace{0.5cm}
    \begin{subfigure}{0.3\textwidth}
        \includegraphics[width=\linewidth]{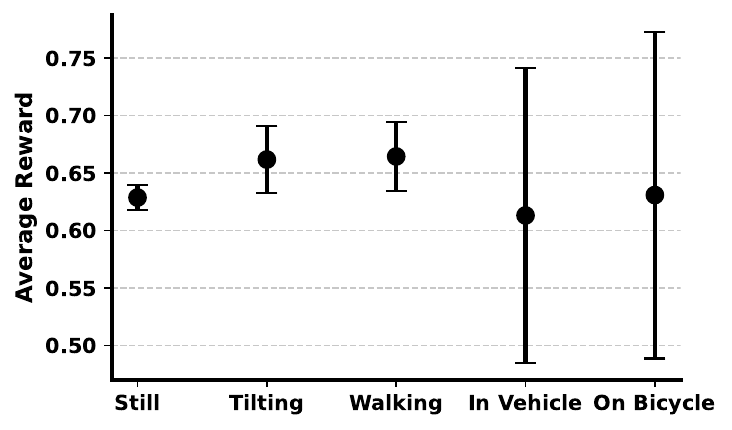}
        \caption{Activity}
        \label{fig:AR_Gapi}
    \end{subfigure}
    \hfill
    \begin{subfigure}{0.3\textwidth}
        \includegraphics[width=\linewidth]{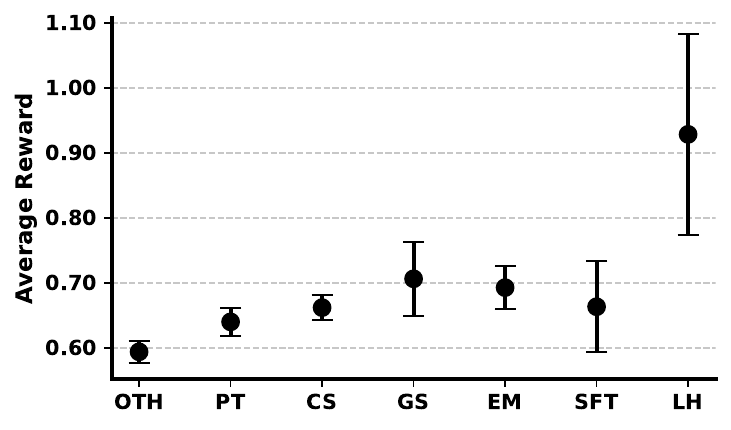}
        \caption{App usage (category)}
        \label{fig:AR_AU}
    \end{subfigure}
    \hfill
    \begin{subfigure}{0.3\textwidth}
        \includegraphics[width=\linewidth]{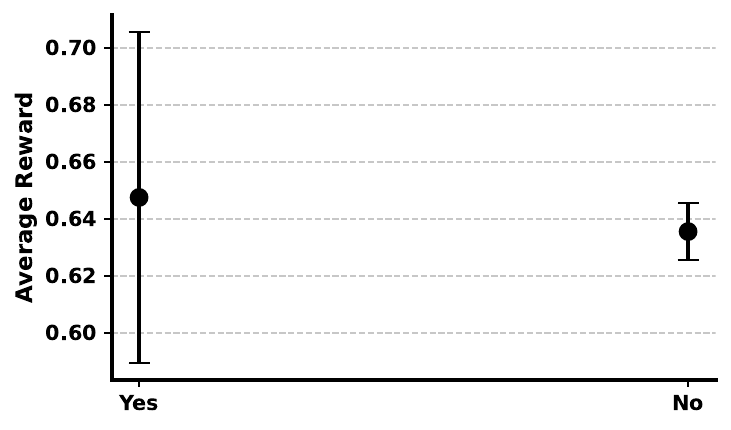}
        \caption{OnCall}
        \label{fig:AR_Calls}
    \end{subfigure}
    \begin{subfigure}{0.3\textwidth}
        \includegraphics[width=\linewidth]{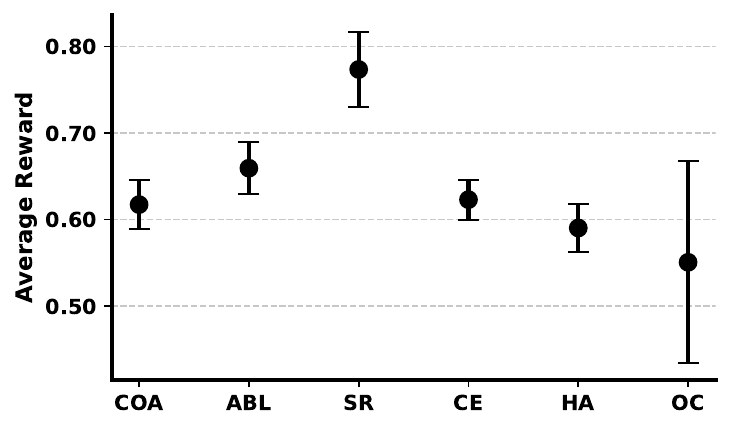}
        \caption{Location (GPS)}
        \label{fig:AR_gps}
    \end{subfigure}
    \Description{}
    \caption{Visual analysis of intervention average rewards under various contextual factors. The abbreviations COA: Campus Open Area, ABL: Academic Building \& Labs, SR: Sports Region, CE: Cafeteria \& Eatery, HA: Hostel Area, OC: Outside Campus, OTH: Other, PT: Productivity \& Tools, CS: Communication \& Social, GS: Games \& Simulation, EM: Entertainment \& Media, SFT: Shopping, Finance \& Travel, LH: Lifestyle \& Health}
    \label{fig:main}
\end{figure}

In our evaluation of participants' location, we found that it had a significant effect on the feasibility of interventions, H(5) = 42.986, p < .001. The post hoc analysis revealed that the sports region significantly differed from the Cafeteria \& Eatery (p = 0.003), Campus Open Area (p = 0.004), and Dormitory Area (p < .001). Upon visualization in Figure \ref{fig:AR_gps}, we observed that participants' feasibility was highest near the sports region. The lowest feasibility was found when participants were outside the campus. This could be because participants are usually outside the campus for leisure, shopping, or other personal activities, making them less willing to engage in interventions. The second highest feasibility was observed in Academic Buildings \& Labs, likely because the designed interventions were simple and quick, taking around 30 seconds to boost mental well-being, making them more acceptable in an academic setting.

\subsection{Effect of active self-reported context and social context on feasibility of intervention (RQ3 \& RQ4)}  \label{section: analysis_3}

\begin{figure}[h]
    \centering
    \begin{subfigure}{0.6\textwidth}
        \includegraphics[width=\linewidth]{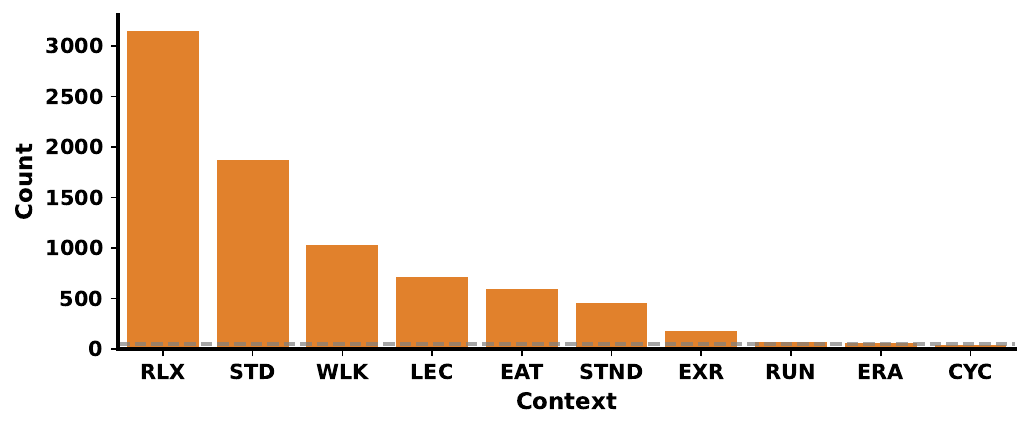}
        \caption{Activity Context}
        \label{fig:AR_UR_C}
    \end{subfigure}
    \hfill
    \begin{subfigure}{0.35\textwidth}
        \includegraphics[width=\linewidth]{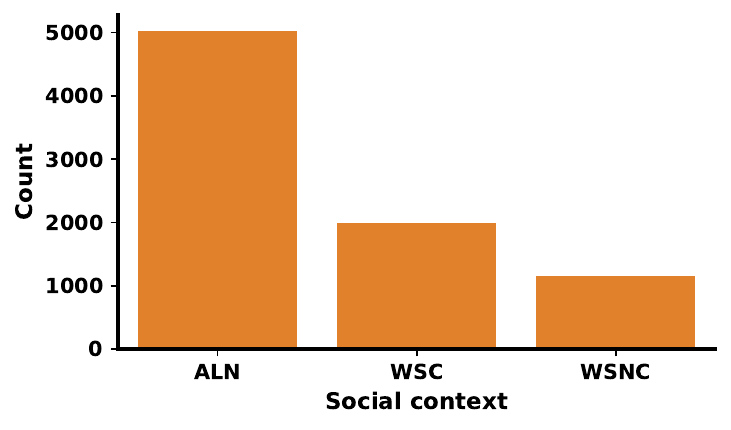}
        \caption{Social context}
        \label{fig:AR_UR_SC}
    \end{subfigure}
    \Description{}
    \caption{Frequency of different (a) activity and (b) social contexts reported by the participants.The abbreviations RLX: Relaxing, EAT: Eating, LEC: Attending Lecture, STD: Studying, WLK: Walking, STND: Standing, EXR: Exercise, ERA: In vehicle, CYC: Cycling, RUN: Running, ALN: Alone, WSC: With Someone (Engaged in Conversation), WSNC: With Someone (Not Engaged in Conversation).}
    \label{fig:main}
\end{figure}

During our data collection, participants self-reported the activity they were engaged in (from 10 predefined activities, see Figure \ref{fig:app_design_interface_1}) before responding to the feasibility of the intervention. They also self-reported their social context—whether they were alone, or with someone (engaged or not engaged in a conversation) (see Figure \ref{fig:app_design_interface_2}). Figure \ref{fig:AR_UR_C} shows the number of instances of different contexts reported by participants, while Figure \ref{fig:AR_UR_SC} shows the number of instances of social contexts reported by participants. In this analysis, we aimed to understand the feasibility of the intervention based on the user-chosen activity (context) and their social context.

We found that the activity chosen by the user had a significant effect on the feasibility of the intervention, H(9) = 138.366, p < .001. Post hoc analysis revealed that attending a lecture significantly differed from eating (p = 0.002), exercising (p = 0.002), and walking (p = 0.003) in terms of average reward. Similarly, eating significantly differed from relaxing (p < .001), standing (p = 0.004), and studying (p < .001). Exercise significantly differed from relaxing (p < .001), standing (p = 0.002), and studying (p < .001). Relaxing significantly differed from studying and walking, and standing significantly differed from walking.
Figure \ref{fig:AR_UR_C} shows the average reward across different user-reported activities. From the figure, we observe that the average reward was highest during exercise, followed by eating, while the lowest was during relaxing. This suggests higher feasibility of suggested interventions during exercise and eating activities and lower feasibility during relaxing activities.
However, for the social context, we did not find a significant effect on the feasibility of the intervention. Figure \ref{fig:AR_UR_SC} shows the mean average reward achieved across different social contexts, where it can be seen that the average rewards were approximately the same across all social contexts.

\begin{figure}[h]
    \centering
    \begin{subfigure}{0.6\textwidth}
        \includegraphics[width=\linewidth]{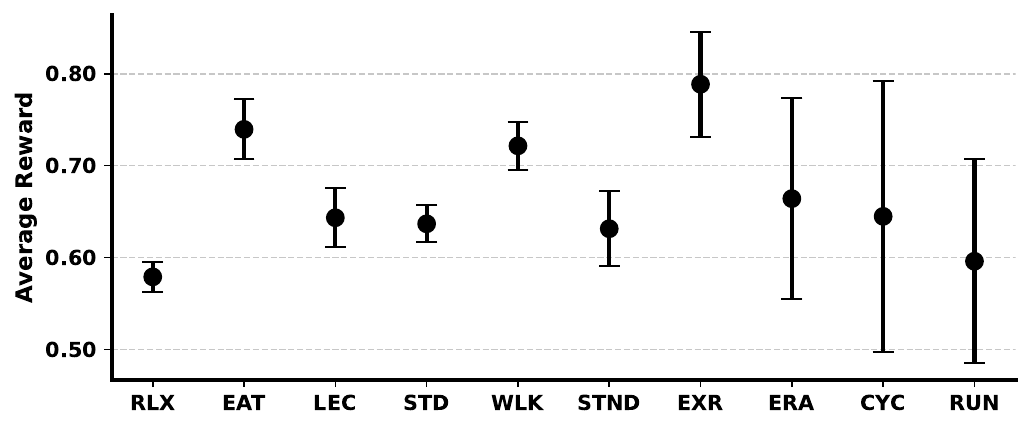}
        \caption{Activity context}
        \label{fig:AR_UR_C}
    \end{subfigure}
    \hfill
    \begin{subfigure}{0.35\textwidth}
        \includegraphics[width=\linewidth]{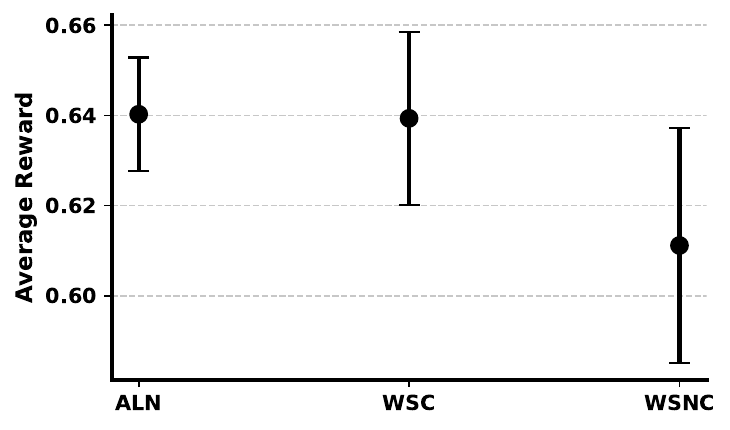}
        \caption{Social context}
        \label{fig:AR_UR_SC}
    \end{subfigure}
    \Description{}
    \caption{Average Reward in different (a) activity, and (b) social contexts. The abbreviations RLX: Relaxing, EAT: Eating, LEC: Attending Lecture, STD: Studying, WLK: Walking, STND: Standing, EXR: Exercise, ERA: In vehicle, CYC: Cycling, RUN: Running, ALN: Alone, WSC: With Someone (Engaged in Conversation), WSNC: With Someone (Not Engaged in Conversation).     }
    \label{fig:main}
\end{figure}

\section{Discussion}
In this section, we discuss the findings and their implications, offering insights for future work. This study aimed to explore the relationship between passively sensed mobile usage and sensor data with the acceptability and feasibility of delivering personalized interventions in real-life settings. We also examined how smartphone data correlates with completion rates and response times to intervention notifications.

Although completion rate and response time are distinct metrics, both help assess how participants respond to notifications. Completion rate reflects the percentage of interventions to which participants responded, while response time captures how quickly they responded. These metrics together provide useful insights for future intervention strategies, particularly regarding when and in what context to send notifications to maximize engagement.

\subsection{Contextual Factors Influencing Intervention Notification Engagement}
Our results suggest that afternoon and evening are the most effective times to deliver intervention notifications to student populations, as these periods show higher completion rates and lower response times. While weekends might seem like suitable days for interventions due to potentially more free time, our findings indicate otherwise. The lack of a structured routine on weekends likely contributes to missed notifications and delayed responses. However, our findings differ from those of Künzler and Mishra et al. \cite{kunzler2019exploring}, who reported that participants were more likely to respond during the evening and on weekends. This difference may be due to the fact that their study involved a general population, whereas our study focused on a student population. Another key finding similar to Künzler and Mishra et al. \cite{kunzler2019exploring} is that when phones are in a charging state and the battery is full, completion rates drop and response times increase. This may be because participants are away from their phones during charging. Similarly, battery levels below 20\% are associated with reduced responsiveness, emphasizing the importance of considering battery conditions before sending interventions.

Device interaction features, such as screen status and lock/unlock events, also showed clear patterns. When the screen is on or the device is unlocked, participants are more likely to engage with notifications, and they tend to respond more quickly \cite{kunzler2019exploring}. This suggests that real-time device interaction can serve as a strong signal for intervention timing. Google's activity recognition also provided valuable insights—walking was associated with the highest notification acceptance and lowest response times \cite{kunzler2019exploring}. On the other hand, the "still" activity was linked to lower engagement, possibly because participants were engaged in other tasks or had their phones set aside.

To further explore device usage behavior, we analyzed app usage categories. We found that participants were more likely to respond quickly to notifications when using communication or social media apps, such as WhatsApp or Instagram. In contrast, they were more likely to ignore notifications when using productivity tools like Google Docs or Google Classroom. This behavior reflects a common pattern where individuals tend to avoid interruptions while focused on work or study tasks. Interestingly, response rates also increased when participants were on a phone call, possibly because they were already actively engaged with their devices.

Location data also provided important contextual cues. Completion rates were higher and response times were lower when participants were near academic buildings. This supports our design choice of sending notifications at the 55th minute of the hour, as class sessions typically run from hh:00 to hh:55, with the following five minutes often used by students to check their phones. Conversely, completion rates were lowest and response times highest when participants were near sports facilities, likely because they were physically active and not carrying their phones. In dining areas such as cafeterias or mess halls, engagement was relatively high, suggesting that participants often use their phones while eating. Although GPS data indicated proximity to sports areas, it could not confirm whether participants were actively engaged in sports, highlighting a current limitation in Human Activity Recognition (HAR) systems. Until HAR methods improve, GPS location near sports facilities can be used as a proxy to avoid sending notifications during those times.

Overall, our findings suggest that various mobile sensor and usage data features can inform the development of intelligent models for timely and context-aware intervention delivery. The key factors identified—time of day, day of the week, battery level and status, screen status, lock/unlock events, recognized activity, app usage category, call state, and location—can be leveraged to train machine learning or reinforcement learning models that optimize notification timing. These models can help maximize participant engagement and the effectiveness of mobile-based health and well-being interventions.

\subsection{User Willingness and Real-World Feasibility of Adaptive Interventions}

To suggest interventions that are more likely to be accepted, we implemented an adaptive intervention module using the Thompson Sampling reinforcement learning algorithm. We implemented separate Thompson Sampling modules for each context (e.g., studying, relaxing, walking) to support context-specific learning of user preferences. Since intervention effectiveness can vary based on the user’s current activity, maintaining independent models for each context ensures more accurate recommendations. Using separate modules avoids confusion during reward updates and prevents biases that could arise in a global model shared across varying intervention sets.

We analyzed the effect of phone sensors and usage patterns on the average reward, which reflects participants’ willingness to perform the suggested tasks and indicates the feasibility of our adaptive intervention approach in real-world settings. Our results showed that interventions delivered in the evening received higher average rewards, suggesting increased willingness at that time—possibly due to lower mood in the evening. Figures \ref{fig:AR_UR_C} and \ref{fig:AR_UR_SC} further show that during weekends, the average reward was lower, possibly because participants were in a better mood and less inclined to accept mental well-being interventions.
Other sensor and usage data did not reveal any strong patterns. However, an interesting observation came from the GPS data: the highest average reward was found when participants were in sports regions. One explanation could be that participants felt physically exhausted and saw the intervention as helpful. The academic location also showed a high average reward, suggesting that our designed interventions were both acceptable and feasible in structured educational settings.

Our analysis of user-reported context also revealed meaningful patterns. The average reward was higher during exercise, indicating that participants were more receptive to well-being interventions during physical activity. Similarly, eating and walking contexts were associated with higher average rewards, suggesting good feasibility for interventions in these situations.
We also examined the role of social context (e.g., being alone or with someone) in influencing intervention acceptance. Surprisingly, there was no significant difference in average reward across different social contexts, indicating that users’ willingness to perform interventions may not be heavily influenced by whether they are alone or accompanied.

\subsection{Limitations and Future work}

Our study was conducted with a student population residing on a university campus, which may limit the generalizability of our findings to broader populations. Additionally, our analysis of intervention acceptability relied solely on passively sensed smartphone data and did not account for potential confounding factors such as age or gender. This decision was based on participants' relatively uniform age range and unequal gender distribution in our sample.

Future work can build on our findings by focusing on university students to design interventions supporting mental well-being. Moreover, our insights on the acceptability of intervention notifications can inform the development of machine learning models that predict optimal moments for delivering such prompts. Contextual factors identified in our study, such as walking, eating, or evening hours, can guide the design of real-world, context-aware interventions that are both acceptable and feasible, ultimately contributing to improved mental health outcomes.

\section{Conclusion}
In this work, we explored the feasibility of adaptive interventions and examined how various passively sensed smartphone data influence user receptivity, measured through completion rate and response time. We conducted a two-week study with 70 student participants, during which they received real interventions to promote mental health well-being. To passively collect smartphone sensor data, actively gather self-reported activity and social context, and deliver interventions, we developed an Android application called LogMe. Our exploratory and inferential analysis found that factors such as time of day, weekday versus weekend, battery level and status, screen interaction, recognized activity, app usage category, and location were significantly associated with completion rate and response time. Additionally, we analyzed the acceptability and feasibility of interventions using the average reward and observed that evening was the most feasible time for intervention delivery. Based on user-reported context, we also found that participants were more willing to perform interventions while walking or eating. These findings suggest that adaptive interventions are feasible and practical in real-world settings.

\bibliographystyle{ACM-Reference-Format}
\bibliography{sample-base}

\end{document}